\documentclass{aa} 

\usepackage{graphicx}
\usepackage{bm}
\usepackage{comment}
\usepackage{gensymb}
\usepackage{arydshln}
\usepackage{booktabs}
\usepackage{multirow}
\usepackage{changepage}

\usepackage[colorlinks=true,citecolor=blue]{hyperref}

\newcommand{\ii}{{{\rm i}}}

\usepackage{txfonts}

\usepackage{natbib}

\begin{document}

\title{Surface measurements of horizontal Reynolds stresses associated with solar inertial modes}
\titlerunning{The Reynolds stress of the solar inertial modes}
\authorrunning{Mandowara et al. 2026}

\author{
Yash~Mandowara \inst{\ref{mps}}\thanks{Corresponding author: \email{mandowara@mps.mpg.de}}
\and
Zhi-Chao~Liang \inst{\ref{mps}}
\and
Richard~S.~Bogart \inst{\ref{stanford}}
\and
Yuto~Bekki \inst{\ref{mps}}
\and
Laurent~Gizon \inst{\ref{mps},\ref{gottingen}}
}

\institute{
Max-Planck-Institut f\"ur Sonnensystemforschung, Justus-von-Liebig-Weg 3, 37077 G\"ottingen, Germany \label{mps}
\and
W.\ W.\ Hansen Experimental Physics Laboratory, Stanford University, Stanford, CA 94305, USA \label{stanford}
\and
Institut f\"ur Astrophysik und Geophysik, Georg-August-Universit\"at G\"ottingen, Friedrich-Hund-Platz 1, 37077 G\"ottingen, Germany \label{gottingen}
}

\date{Received <-->; accepted <-->}
 
\abstract
{
Whether solar inertial modes play a dynamical role in the Sun's convection zone is a topic of growing interest.}
{
We aim to measure the horizontal Reynolds stresses near the solar surface associated with the inertial modes, $\langle u_\theta' u_\phi' \rangle$, where $u_\theta'$ and $u_\phi'$ are the colatitudinal and longitudinal velocity components of each mode.
We consider retrograde propagating equatorial Rossby (ER) modes and high-latitude (HL) inertial modes.
}
{We used 15 years of horizontal flow maps near the Sun's surface obtained from $5\degree$-tile ring-diagram analysis applied to HMI/SDO Doppler maps.
Fourier analysis, frequency filtering, and singular value decomposition were used to extract the mode profiles of the 16 strongest known inertial modes. 
}
{
Among these modes, the $m=1$ HL mode at frequency $-89.1$~nHz in the Carrington frame contributes most strongly to the total Reynolds stress, reaching $44$~m$^2$s$^{-2}$ near latitude $75\degree$ during solar minimum, above the latitude of its critical layer.
In contrast, the ER modes fall into two regimes: the Reynolds stress of modes with $3\leq m\leq5$ peaks above their critical layers, whereas that of modes with $m\geq6$ switches sign around their critical layers.
The cumulative Reynolds stress from all the ER modes reaches $2$~m$^2$s$^{-2}$ near a latitude of $\sim 8\degree$ during solar maximum.}
{The Reynolds stresses associated with the inertial modes oppose those produced by turbulent convection and active-region inflows. 
While the surface Reynolds stresses from the ER modes are much weaker than those of the convective flows, those from the HL modes are comparable to or even exceed the Reynolds stresses from turbulent convection during solar minimum, indicating that the HL modes may play an important dynamical role in the transport of angular momentum in the polar regions.}

\keywords{Sun: interior -- Sun: helioseismology -- Sun: oscillations}

\maketitle

\section{Introduction}

The Sun's convection zone exhibits strong latitudinal differential rotation, with the equator rotating faster than the poles \citep[e.g.,][]{schou1998}. 
This differential rotation is believed to be sustained largely by Reynolds stress of rotating turbulent convection \citep[e.g.,][]{brun2002,miesch2008} or by Maxwell stress of magnetic fields \citep[e.g.,][]{hotta2022}.
Identifying the physical mechanisms responsible for the  angular momentum transport remains an important challenge in solar physics.

In addition to turbulent magneto-convection, global-scale waves may also contribute to angular momentum transport. 
In rotating fluids, low-frequency modes of global oscillations called inertial modes exist.
They are restored by the Coriolis force and hence have frequencies on the order of the rotation rate \citep{greenspan1968}.
Recently, using the long-term observations of solar near-surface flows, \citet{loeptien2018} reported the unambiguous detection of the equatorial Rossby (ER) modes, a particular subclass of inertial modes whose flow motions are quasi-toroidal. 
These Rossby modes propagate in the retrograde direction in a frame rotating with the Sun and have typical velocity amplitudes of about $1$--$2$~m~s$^{-1}$.
Subsequent studies further revealed that the spectrum of the Sun is richer than previously thought. 
\citet{gizon2021} reported the modes at middle and high latitudes that owe their existence to differential rotation.
In particular, the high-latitude (HL) mode with azimuthal order $m=1$ possesses the largest velocity amplitude of $10$--$20$~m~s$^{-1}$ among all observed modes \citep[][]{liang2025}.
The large amplitude of this mode likely results from fluid instabilities inside the convection zone \citep[][]{bekki2022a,fournier2022}.
Furthermore, \citet{hanson2022} reported an additional family of modes called high-frequency retrograde (HFR) modes, with weak amplitudes ($\lesssim 0.5$~m~s$^{-1}$).
These HFR modes are characterized by substantial horizontal divergences, implying that they are strongly non-toroidal in nature \citep[][]{hanson2025}.
Despite these advances, the dynamical role of solar inertial modes in the convection zone remains observationally unconstrained.

Solar inertial modes may influence the large-scale dynamics of the convection zone in several ways. 
For example, HL inertial modes can become baroclinically unstable and modify the background pole-to-equator temperature difference by transporting heat equatorward, thereby affecting the solar differential rotation \citep[][]{bekki2024}.
Another possible dynamical effect arises from the Reynolds stresses associated with these modes, which can transport angular momentum.
In fact, theoretical studies suggest that, in the presence of differential rotation, ER modes are strongly distorted in the viscous critical layers formed by corotation resonance, leading to substantial Reynolds stresses associated with equatorward angular momentum transport \citep[][]{gizon2020b,bekki2022a}.  
Measuring these Reynolds stresses provides a direct way to assess whether inertial modes contribute significantly to the angular momentum balance.

The flows ($\mathbf{u}(r,\theta,\phi,t)$) in the solar convection zone vary with depth ($r$), colatitude ($\theta$), longitude ($\phi$), and time ($t$), and can be decomposed as 
\begin{align}
&\mathbf{u}\; = \mathbf{u}_\mathrm{mer} + r\sin\theta~\Omega~{\hat\phi} + \mathbf{u}',\\
&\mathbf{u}' = \mathbf{u}_{\rm conv} + \mathbf{u}_{\rm AR} + \mathbf{u}_{\rm modes},
\label{eq:flow_comp}
\end{align}
where $\mathbf{u}_\mathrm{mer}(\theta)$ is the meridional circulation, $\Omega(r,\theta,t)$ is the differential rotation of the Sun, $\mathbf{u}_{\rm conv}(r,\theta,\phi,t)$ is turbulent convection, $\mathbf{u}_{\rm AR}(r,\theta,\phi,t)$ represents active-region flows, and $\mathbf{u}_{\rm modes}(r,\theta,\phi,t)$ are the low-frequency inertial modes.
Non-axisymmetric flows, represented by $\mathbf{u}'$, constitute the primary data in this study.
The angular momentum conservation equation \citep{ruediger89},
\begin{align}
\rho r^2\sin^2\theta\frac{\partial \Omega}{\partial t}
    &= -\mathbf{\nabla} \cdot \left[ \rho r^2\sin^2\theta\Omega\mathbf{u}_\mathrm{mer} + \rho r\sin\theta\langle\mathbf{u}'u_\phi'\rangle_\phi \right] \nonumber \\
    &\quad + \mbox{magnetic effects},
\label{eq:amb}
\end{align}
where $\rho(r)$ is the radially varying background density and $\langle\cdot\rangle_\phi$ denotes a longitudinal average.
This equation relates changes in the rotation rate to the advection by meridional circulation, Reynolds stresses (including turbulent diffusion), and magnetic effects, the latter of which are beyond the scope of the present study. 
The Reynolds stresses can be decomposed into contributions from the different flow components in Eq.~\eqref{eq:flow_comp}.
This study focuses on the horizontal Reynolds stress $Q$, defined as the longitudinal mean of the product of the two horizontal velocity components.

We present observational measurements of the horizontal Reynolds stresses associated with solar inertial modes. 
Using $15$ years of horizontal flow maps of the Sun's near-surface layer, we extract the velocity eigenfunctions of the inertial modes with strong enough signal-to-noise ratio and compute their associated Reynolds stresses. 
We investigate their spatial structure and temporal variability and assess whether their combined contribution is significant in the context of Reynolds stresses from non-axisymmetric flows. This can in turn tell us if they are a significant component of the angular momentum balance.

\begin{figure*}
\centering
\sidecaption
\includegraphics[width=12cm]{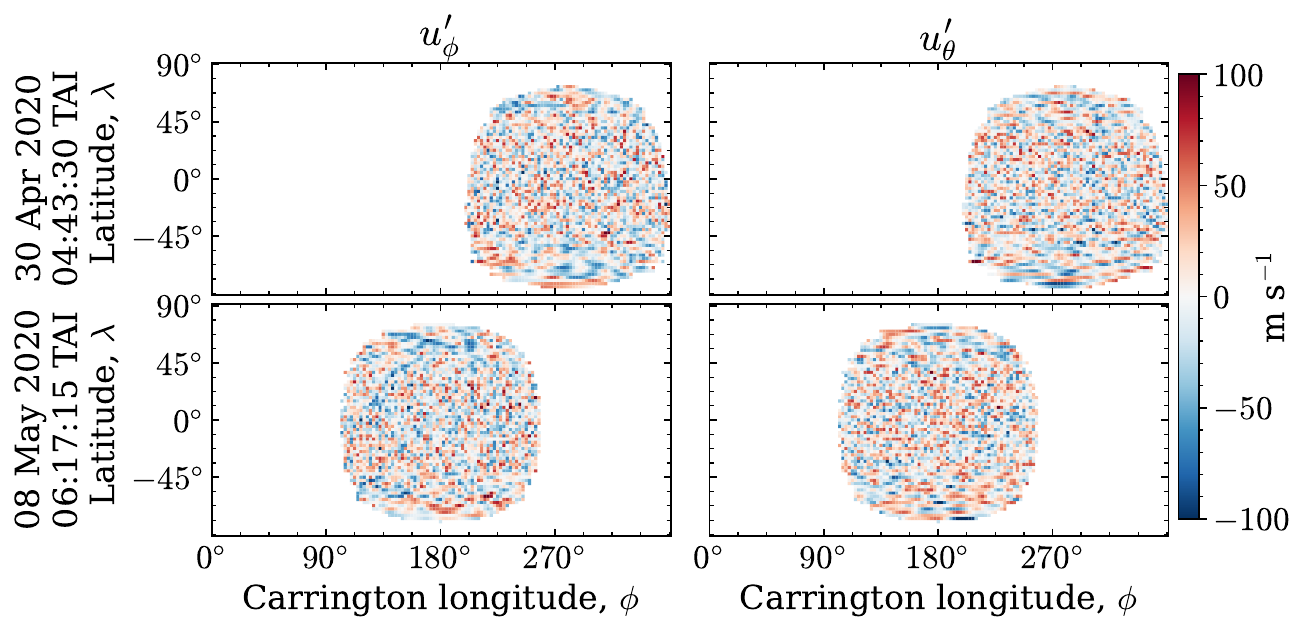}
\caption{
Example horizontal velocity maps from the HMI $5\degree$-tile ring-diagram analysis in Carrington frame. 
Left and right panels show the prograde velocity $u'_\phi$ and the southward velocity $u'_\theta$, respectively, at a depth of $0.997~R_\odot$.
Each ring-diagram snapshot corresponds to the average flow over a duration of 9 hours, whose midpoint is shown in the label.
In order to remove mean flows and $B_0$-angle related systematics, we remove the temporal mean and yearly variations from the data in Stonyhurst coordinates.
}
\label{f:flow maps}
\end{figure*}

\section{Data analysis}

\subsection{Overview of the dataset}
\label{sec:ov}

For this study, we used data from the Helioseismic and Magnetic Imager \citep[HMI;][]{scherrer2012,schou2012} on board the Solar Dynamics Observatory \citep[SDO;][]{pesnell2012}. 
Specifically, we used horizontal velocity maps produced by the HMI ring-diagram pipeline \citep{bogart2011a, bogart2011b, bogart2015}. We used the $5\degree$-tile ring-diagram flow maps from the data series \texttt{hmi.rdubinserD\_fd05}, available through the Joint Science Operations Center (JSOC). In this method, full-disk Dopplergrams of the Sun are tracked to follow a $5\degree\times5\degree$ patch of the solar photosphere over 1/72 of a Carrington rotation. These tracked surface regions are Fourier transformed to yield 3D power spectra.
These power spectra contain acoustic (p-mode) and surface gravity (f-mode) oscillations Doppler shifted by subsurface flows.
These shifts are fitted using a model \citep{basu1999} to obtain flow velocities.

A smaller number of  modes can be reliably fitted in the $5\degree$-tile power spectra than in the traditional $15\degree$-tile power spectra \citep{hill1988, basu1999}.
The pipeline rejects outlying fits: at a given Stonyhurst location, a fitted flow value at a particular frequency and radial order is discarded if it differs by more than $75$~m~s$^{-1}$ from the median of that same quantity, taken over all samples within one year. A flow value is then reported for a given flow component and radial order only if at least 15 fits survive this rejection. Components and radial orders that fail this criterion are treated as missing data (Sect.~\ref{sec:dpp}).

Each radial order is assigned a depth equal to the mean lower turning point of its modes, computed from a standard solar model. The dataset therefore has four points in depth, corresponding to the p-modes with radial orders $n = 1, 2, 3$ and to the f-modes. This dataset differs from the one used in the $5\degree$-tile analysis of \citet{gizon2021}, where all radial orders are averaged to obtain a single mean flow per tile. For this study, we only consider the f-mode data, which probe the top $2$~Mm below the solar surface ($r=0.997~R_\odot$).

The key velocity products provided are $(u_\theta, u_\phi)$, where $u_\theta$ is southward and $u_\phi$ is prograde. 
These horizontal flows are given in units of m~s$^{-1}$ as functions of colatitude ($\theta$), longitude ($\phi$), and time ($t$).
For convenience, we also use the latitude defined as $\lambda = 90\degree - \theta$.
The maximum available latitude varies between $\pm72.5\degree$ and $\pm 85\degree$, depending on the solar $B_0$ angle.
The maximum extent in Stonyhurst longitude is $\pm 77.5\degree$ near the equator and decreases toward higher latitudes. 
The $5\degree$-tile data have a 50 \% overlap with neighboring tiles, giving us the nominal resolution of $\Delta \phi=\Delta \theta=2.5\degree$, for tiles at and below $40\degree$ latitude.
At higher latitudes, the nominal resolution of $\Delta \theta=2.5\degree$ is maintained, but there are substantial differences in longitude. 
The resolution in longitude, $\Delta\phi$ is $5\degree$ for latitudes ranging $42.5\degree$ to $65\degree$, $7.5\degree$ for latitudes from $67.5\degree$ to $72.5\degree$, and $10\degree$ for latitudes from $75\degree$ to $85\degree$. 

The dataset has $N_t=14666$ points in time from May 2010 to June 2025, with a temporal resolution, $\Delta t= 9.09$~hours. 
We note that $N_t$ includes 216 empty frames (1.5\% of the total) for which no flow maps are available.

\subsection{Data post-processing}\label{sec:dpp}

The following post-processing closely follows the procedure described in \cite{proxauf2020}, with some modifications to account for the different grid and tile size. 
We discard all points in the flow maps with east-west or north-south velocities above 300~m~s$^{-1}$.
The rejection rate for this step is $0.0332$~\%. 

Following this, we remove the yearly systematics and mean flows by fitting each pixel of data in time with a sinusoidal function in time with a constant offset. 
This constant offset comprises mean flows in the Sun, along with constant systematics such as center-to-limb effects.
On the other hand, the yearly systematics arise from variations in the solar $B_0$ angle.
The fit performed using least squares is then subtracted from the data.
The fit is performed using only the available data points.
This removes the largest systematics such as center-to-limb effects, and the zeroth order of mean flows like differential rotation and meridional circulation.

Data are missing for five reasons:
(i) too few modes were resolved for a reliable fit;
(ii) outlier rejection;
(iii) interruptions in the observations, leaving whole frames empty;
(iv) the coarser longitudinal spacing of tiles at high latitudes (Sect.~\ref{sec:ov}); and
(v) incomplete coverage of the Sun.
We fill gaps (i)--(iv) by interpolating onto a uniform grid in longitude ($\Delta\phi = 2.5\degree$) and time.
Gaps (v) cannot be filled; we correct for them using $\eta(\theta)$ (Sect.~\ref{symm}).
The flow maps are then transformed to the Carrington frame ($N_\phi = 144$ over $360\degree$). Figure~\ref{f:flow maps} shows an example, in which the unfilled gaps (v) are visible.

\subsection{Fourier transform of velocity maps}
\label{symm}

The post-processed velocity, $\mathbf{u}'=(u_\theta', u_\phi')$, is the non-axisymmetric component of horizontal flows used in Eq.~\eqref{eq:flow_comp}.
We then decompose each velocity component into parts that are symmetric ($+$) and antisymmetric ($-$) across the equator:
\begin{equation}
u^\pm_j(\theta, \phi, t) = [u'_j(\theta, \phi, t) \pm u'_j(\pi - \theta, \phi, t)]/2.
\label{eq:symm}
\end{equation}
Where only one of $u'_j(\theta)$ and $u'_j(\pi-\theta)$ is available, we use that value alone instead of averaging it with a missing value, which maximizes coverage at high latitudes.

Observed inertial modes fall into two classes, according to the north-south symmetry of their radial vorticity $\zeta_r$. 
Modes with symmetric radial vorticity ($\zeta_r^+$) have a symmetric latitudinal velocity and an antisymmetric longitudinal velocity, $(u_\theta^+, u_\phi^-)$. 
Modes with antisymmetric radial vorticity ($\zeta_r^-$) have the opposite combination, $(u_\theta^-, u_\phi^+)$.

We perform the 2D discrete Fourier transform in time and longitude:
\begin{equation}\label{eq:FT2D}
U_j^\pm(\theta,m,\nu) = \frac{1}{\eta(\theta)}\frac{1}{N_t}\sum_{t}\frac{1}{N_\phi}\sum_{\phi} u_j^\pm(\theta,\phi, t)\, e^{-\ii m\phi + 2\ii\pi\nu t},
\end{equation}
where $m$ is the azimuthal order and $\nu$ is the frequency, with a resolution of $2.08$~nHz.
The factor $\eta(\theta)$ is the fraction of available data points at each latitude, and its reciprocal corrects for the amplitude loss due to missing data.
The correction $1/\eta$ (Fig.~\ref{f:ac}) is nearly constant below $60\degree$ and rises steeply above.

\begin{figure}
\centering
\includegraphics[width=0.8\hsize]{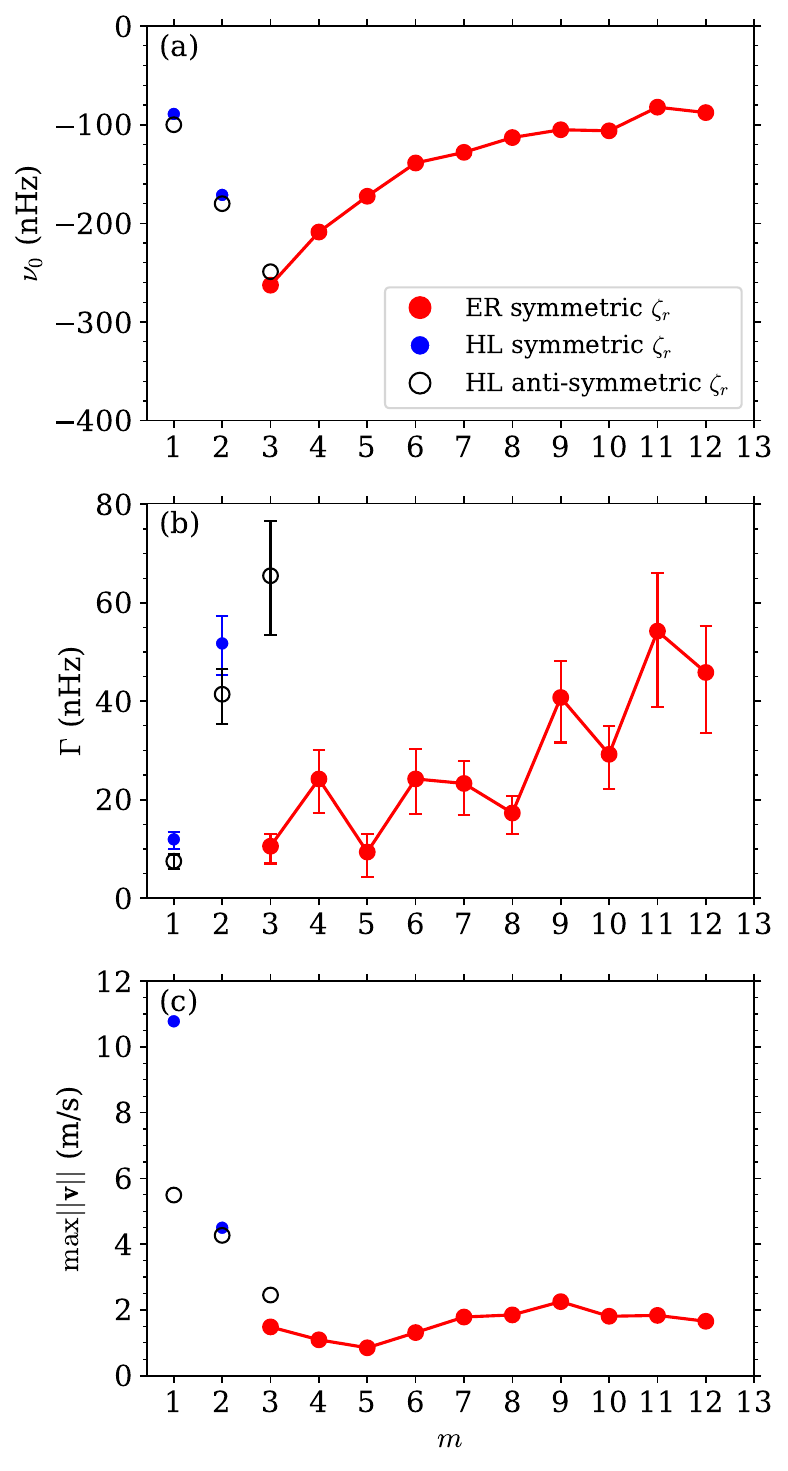}
\caption{
Frequencies (a), linewidths (b), and amplitudes (c) of the 16 inertial modes listed in Table~\ref{t:mode params}.
Red filled circles denote the ER modes, all of which have north-south symmetric radial vorticity.
Blue filled circles denote the HL modes with symmetric radial vorticity, while black open circles denote the HL modes with antisymmetric radial vorticity.
The amplitude is defined as the maximum velocity of the 2D surface profile, $\mathbf{v}(\theta,\phi)$, of each mode.
The frequency uncertainties are smaller than the symbol size.
The uncertainties in $\max\|\mathbf{v}\|$ are smaller than 0.3~m~s$^{-1}$ for the ER modes and 1~m~s$^{-1}$ for the HL modes.
}
\label{f:mode_parameters}
\end{figure}

\subsection{Measuring mode frequencies and linewidths}\label{s:flw}

Conventionally in the study of solar inertial modes, we focus on positive $m$, and retrograde modes are identified by negative frequencies.
In this study, we focus on inertial modes with $m \leq 13$ and $|\nu|<\Omega_{\rm Carr}/2\pi$ in the Carrington frame. 
The analyzed modes include the high-latitude (HL) modes with $1 \le m \le 3$, which may exist in both north-south symmetries, and the equatorial Rossby (ER) modes with $3 \le m \le 13$, which are all symmetric in $\zeta_r$.
Although excess power associated with ER modes is visible in the power spectrum of $U_\theta$ up to $m=17$, we restrict our analysis to $m \le 13$, as mode extraction becomes very difficult at these $m$ due to noise and increasing linewidths of ER modes.

For each mode with a given fixed value of $m$, we generate a 1D power spectrum at that azimuthal order averaged over a specific range in latitude, 
\begin{equation}
P^\pm_{j}(\nu) = \left\langle |U^\pm_j(\theta,m,\nu)|^2\right\rangle_\theta,~\mbox{for a fixed}~m,
\end{equation}
where $\langle\cdot\rangle_\theta$ denotes a latitudinal average.
The power is averaged over latitudes $60\degree$ -- $75\degree$ for the HL modes, and $0\degree$ -- $15\degree$ for the ER modes. 
We estimate important mode parameters such as the frequency $\nu_0$, linewidth $\Gamma$, and amplitude $A$ by fitting a Lorentzian profile to the latitudinally averaged power spectrum $P^\pm_j(\nu)$. 
The Lorentzian profile used for the fit is given by
\begin{equation}
L(\nu,\mathbf{h}) = \frac{A}{1+[(\nu-\nu_0)/(\Gamma/2)]^2} + B, \label{eq:lorentz} 
\end{equation}
where $\mathbf{h}$ is the set of parameters describing a mode: $\mathbf{h} = (\nu_0, \Gamma, A, B)$ and $B$ is the background.
The fit is performed using a maximum likelihood estimator by minimizing
\begin{equation}
\sum_{\nu} \left(\ln[L(\nu,\mathbf{h})] + P^\pm_j(\nu)/L(\nu,\mathbf{h})\right),\label{eq:mle}
\end{equation}
where the fitting range in $\nu$ is carefully chosen for each mode to ensure a smooth background. 
Table~\ref{t:mode params} lists important fitted parameters such as mode frequency and linewidth for 16 inertial modes.
Uncertainties in frequency and linewidth were estimated using a Monte Carlo method. We generated 1000 realizations of the fitted profile perturbed according to the $\chi^2$-errors, performed the fit for each realization, and took the standard deviation across all 1000 realizations as the reported uncertainty.

\begin{table*}
\centering
\renewcommand{\arraystretch}{1.3}
\caption{Solar inertial modes analyzed in this study and their basic parameters obtained using $5\degree$-tile HMI ring-diagram data for 2010--2025.}
\begin{tabular}{cccccccccc}
\noalign{\vskip -9pt}
\hline
\hline
\noalign{\vskip -1pt}
$m$ & Mode & Frequency  & Linewidth & Latitude of & $\max\|\mathbf{v}\|$ & Latitude & $\max Q$ & $\max Q$ & $\max Q$ \\
&  & $\nu_0$ (nHz) & $\Gamma$ (nHz) & critical layer & (m~s$^{-1}$) & at $\max Q$ & (m$^2$~s$^{-2}$) & (m$^2$~s$^{-2}$) & (m$^2$~s$^{-2}$) \\
&&&&$\lambda_c$ at $r=R_\odot$&&2010--25&2010--25& 2021--25 & 2016--20 \\
\noalign{\vskip 1.5pt}
\hline
\noalign{\vskip 1.5pt}
1 & HL$+$ & $-89.1^{+0.8}_{-0.8}$ & $11.9^{+1.5}_{-2.0}$ & 58.1$^\circ$ & 11.6 & 77.5$^\circ$ & 27.7 & 12.4 & 43.6 \\
1 & HL$-$ & $-100.0$ & $-$ & 63.8$^\circ$ & 5.6 & 77.5$^\circ$ & 4.5 & 8.3 & 1.9 \\
2 & HL$+$ & $-171.2^{+2.4}_{-2.3}$ & $51.7^{+5.7}_{-6.3}$ & 58.1$^\circ$ & 5.3 & 72.5$^\circ$ & 6.7 & 5.1 & 8.4 \\
2 & HL$-$ & $-180.0^{+1.9}_{-2.3}$ & $41.4^{+5.1}_{-6.0}$ & 58.1$^\circ$ & 5.4 & 72.5$^\circ$ & 7.0 & 5.9 & 9.8 \\
3 & HL$-$ & $-249.0^{+3.8}_{-3.5}$ & $65.5^{+11.0}_{-12.1}$ & 56.2$^\circ$ & 3.4 & 77.5$^\circ$ & 2.8 & 5.5 & 3.1 \\
\noalign{\vskip 1.5pt}
\hdashline[1.5pt/2.5pt]
\noalign{\vskip 1.5pt}
3 & ER$+$ & $-262.7^{+1.3}_{-1.2}$ & $10.5^{+2.6}_{-3.5}$ & 58.1$^\circ$ & 1.5 & 70.0$^\circ$ & 0.4 & 0.3 & 1.3 \\
4 & ER$+$ & $-208.8^{+2.3}_{-2.4}$ & $24.2^{+5.9}_{-6.9}$ & 45.0$^\circ$ & 1.7 & 57.5$^\circ$ & 0.5 & 0.5 & 0.6 \\
5 & ER$+$ & $-172.5^{+1.8}_{-1.7}$ & $9.4^{+3.6}_{-5.1}$ & 37.5$^\circ$ & 0.8 & 42.5$^\circ$ & 0.1 & 0.2 & 0.3 \\
6 & ER$+$ & $-138.8^{+2.6}_{-2.7}$ & $24.2^{+6.1}_{-7.2}$ & 30.0$^\circ$ & 1.3 & 17.5$^\circ$ & 0.1 & 0.4 & 0.3 \\
7 & ER$+$ & $-128.0^{+2.1}_{-2.0}$ & $23.3^{+4.5}_{-6.4}$ & 26.2$^\circ$ & 1.8 & 10.0$^\circ$ & 0.2 & 0.4 & 0.2 \\
8 & ER$+$ & $-113.0^{+1.5}_{-1.6}$ & $17.3^{+3.4}_{-4.2}$ & 22.5$^\circ$ & 1.8 & 7.5$^\circ$ & 0.2 & 0.2 & 0.2 \\
9 & ER$+$ & $-105.1^{+2.7}_{-2.7}$ & $40.8^{+7.5}_{-9.1}$ & 20.6$^\circ$ & 2.2 & 7.5$^\circ$ & 0.3 & 0.4 & 0.1 \\
10 & ER$+$ & $-106.2^{+2.3}_{-2.1}$ & $29.2^{+5.8}_{-7.0}$ & 18.8$^\circ$ & 1.8 & 7.5$^\circ$ & 0.2 & 0.2 & 0.2 \\
11 & ER$+$ & $-82.2^{+4.2}_{-4.1}$ & $54.2^{+11.9}_{-15.4}$ & 15.0$^\circ$ & 1.8 & 5.0$^\circ$ & 0.2 & 0.3 & 0.1 \\
12 & ER$+$ & $-87.7^{+3.5}_{-3.5}$ & $45.8^{+9.5}_{-12.3}$ & 15.0$^\circ$ & 1.7 & 7.5$^\circ$ & 0.1 & 0.2 & 0.2 \\
13 & ER$+$ & $-73.9^{+4.3}_{-4.6}$ & $48.8^{+11.5}_{-16.1}$ & 11.2$^\circ$ & 1.7 & 7.5$^\circ$ & 0.2 & 0.3 & 0.2 \\
\bottomrule
\noalign{\vskip -5pt}
\end{tabular}
\tablefoot{
Frequencies are reported in the Carrington frame. 
`HL' and `ER' denote the high-latitude and equatorial Rossby modes, respectively.
The symbols `$+$' and `$-$' denote the north-south symmetry and antisymmetry, respectively, of the radial vorticity $\zeta_r$.
For the $m=1$ HL$-$ mode, no fitting was performed; instead, a filtering range of $\nu_0\pm15$~nHz was used.
The period from April 2021 to April 2025 corresponds to enhanced solar activity.
The period from April 2016 to April 2020 corresponds to comparatively low solar activity.
}
\label{t:mode params}
\end{table*}

\subsection{Eigenfunction extraction and mode Reynolds stress}\label{s:fee}

We extract the horizontal velocity eigenfunctions using a singular value decomposition (SVD), following the approach used by \citet{proxauf2020} and \citet{gizon2021}. 
For each mode of azimuthal order $m$, we apply a rectangular bandpass filter, given by
\begin{equation}
F(\nu) = \mbox{rect}\left(\frac{\nu - \nu_0} {4\Gamma}\right)= 
\begin{cases}
    1, & |\nu-\nu_0|\leq 2\Gamma,\\
    0, & \text{otherwise.}
\end{cases}    
\end{equation}
The frequency-filtered velocity data are transformed back to the time domain. For each mode with $m>0$, the filtered velocity $\mathbf{u}_0$ is obtained from $\mathbf{U}=[U_\theta, U_\phi]$ as
\begin{equation}
\mathbf{u}_0^\pm(\theta, t) = \sum_{\nu} \mathbf{U}^\pm(\theta, \nu)\  F(\nu) \exp({-2\ii\pi\nu t}).
\end{equation}
Only data within $|\lambda|\leq 60\degree$ and $|\lambda|\leq 80\degree$ are included in the SVD for the ER and HL modes, respectively.
We then perform an SVD of the filtered velocity to separate the spatial and temporal dependence of the modes:
\begin{equation}
\mathbf{u}_0^\pm(\theta, t) = \sum_{q=1}^{2N_\theta}\sigma_q T_q(t) \mathbf{X}_q^*(\theta),
\end{equation}
where $\sigma_q$ are the non-negative singular values, arranged in decreasing order, and $N_\theta=73$ is the number of grid points in colatitude. The temporal and colatitudinal singular vectors are $T_q(t)$ and $\mathbf{X}_q(\theta)$, respectively.

We assume that the dominant signal in the filtered velocity is the mode of interest. Accordingly, the largest singular value $\sigma_1$ and its associated temporal dependence $T_1(t)$ correspond to the target mode.
The filtered complex eigenfunction $\mathbf{v}$, defined at all latitudes, is obtained by projecting the filtered velocity $\mathbf{u}_0^\pm$ onto the leading temporal singular vector $T_1(t)$:
\begin{equation}
\mathbf{v}(\theta) = \sum_t T_1^*(t) \mathbf{u}_0^\pm(\theta,t).
\label{eq:svd_projection}
\end{equation}

To further reduce the noise, we smooth the 1D eigenfunctions by expanding them in a basis of Legendre polynomials. 
The eigenfunctions $v_j(\theta)$ are reconstructed by truncating the expansion at a cut-off degree $\ell_c=25$ (see Appendix~\ref{s:err}).
Finally, for each mode, the 2D eigenfunction on the sphere is given by
\begin{equation}
v_j(\theta,\phi) = 2\Re [v_j(\theta) \exp({\ii m\phi})].
\end{equation}
From these eigenfunctions, we compute the horizontal Reynolds stress $Q(\theta)$ of each mode as the longitudinal average of the product of the two horizontal velocity components,
\begin{equation}
Q(\theta) = \langle v_\theta(\theta, \phi) v_\phi(\theta, \phi)\rangle_\phi\;.
\label{eq:RS}
\end{equation}

\begin{figure*}[h!]
\sidecaption
\includegraphics[width=12cm]{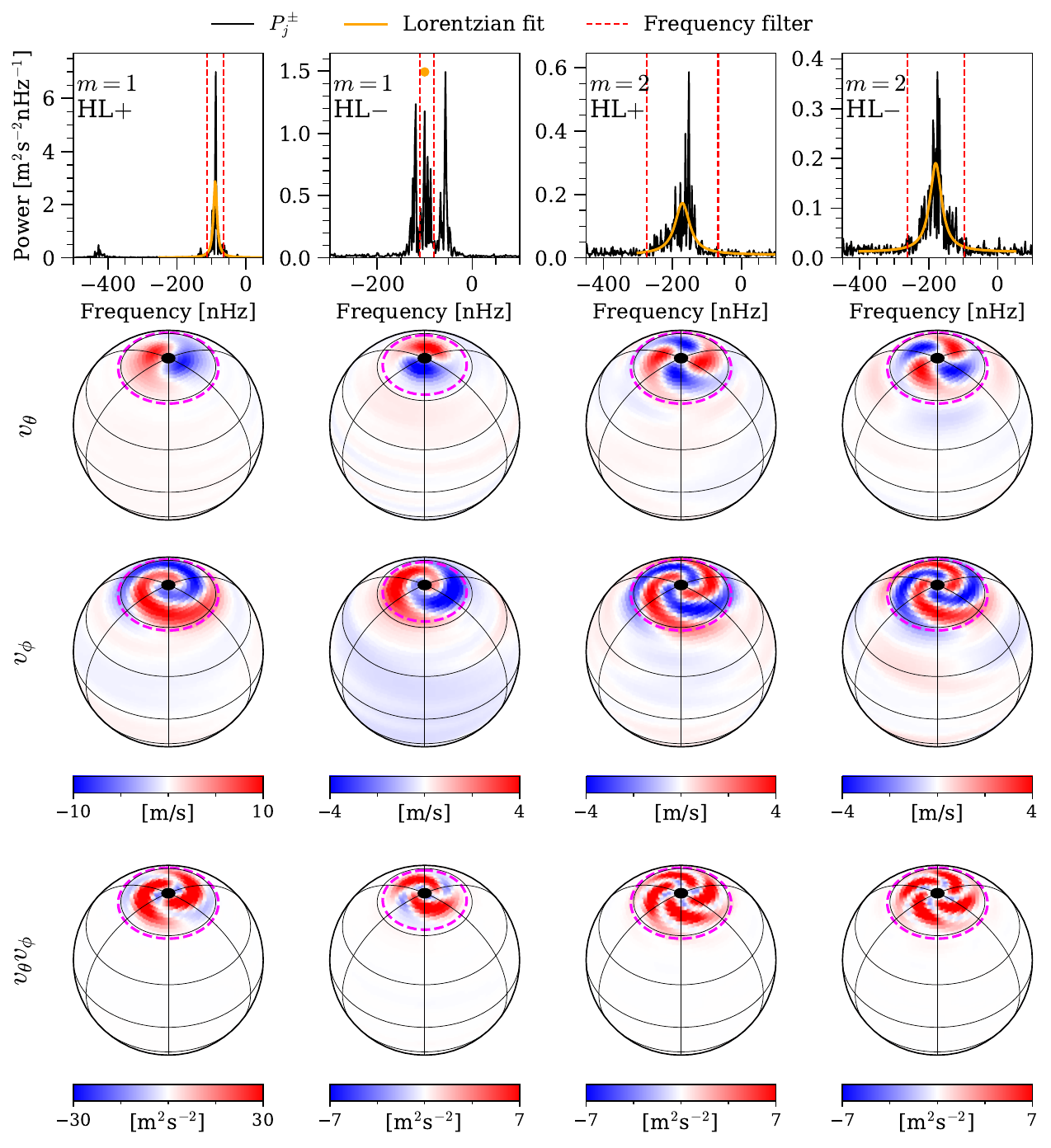}
\caption{
Observed velocity eigenfunctions of the $m=1$ and $m=2$ HL modes listed in Table~\ref{t:mode params}.
These are the HL modes with the largest amplitudes.
(\textit{Top}) Latitudinally averaged velocity power spectra $P^{\pm}_{j}$. The HL$+$ modes use $P^-_\phi$, the HL$-$ modes use $P^+_\phi$, and the ER$+$ modes use $P^+_\theta$ for the Lorentzian fits.
The yellow curves show the fitted Lorentzian profiles, and the dashed red lines indicate the frequency ranges used for the SVD to extract the eigenfunctions. 
(\textit{Middle 2 panels}) Surface eigenfunctions of the southward velocity, $v_{\theta}$, and the prograde velocity, $v_{\phi}$.
The black regions near the poles indicate where no data are available.
The dashed magenta line marks $\lambda_c$ at the solar surface.
(\textit{Bottom}) Products of the two horizontal velocity eigenfunctions, $v_{\theta}v_{\phi}$.
For the $m=1$ HL$-$ mode, no Lorentzian fitting was performed; instead, a filtering range of $-100\pm15$~nHz was used.
}
\label{f:ef1-4}
\end{figure*} 

\begin{figure*}
\sidecaption
\includegraphics[width=12cm]{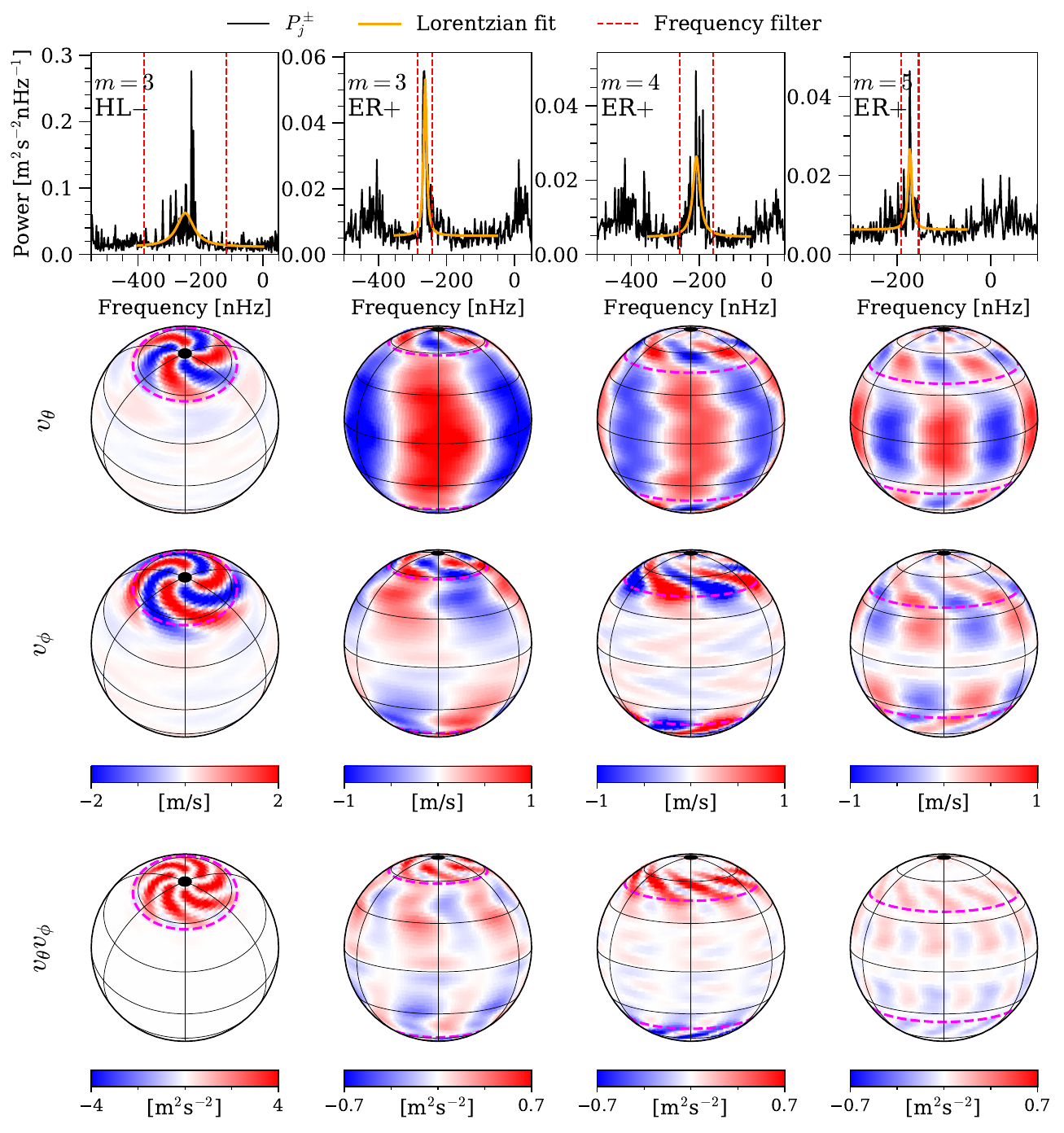}
\caption{
Same as Fig.~\ref{f:ef1-4}, but for modes with $3\leq m\leq 5$.
The leftmost mode is a HL mode, while the rest are ER modes. 
}
\label{f:ef5-8}
\end{figure*}

\begin{figure*}
\sidecaption
\includegraphics[width=12cm]{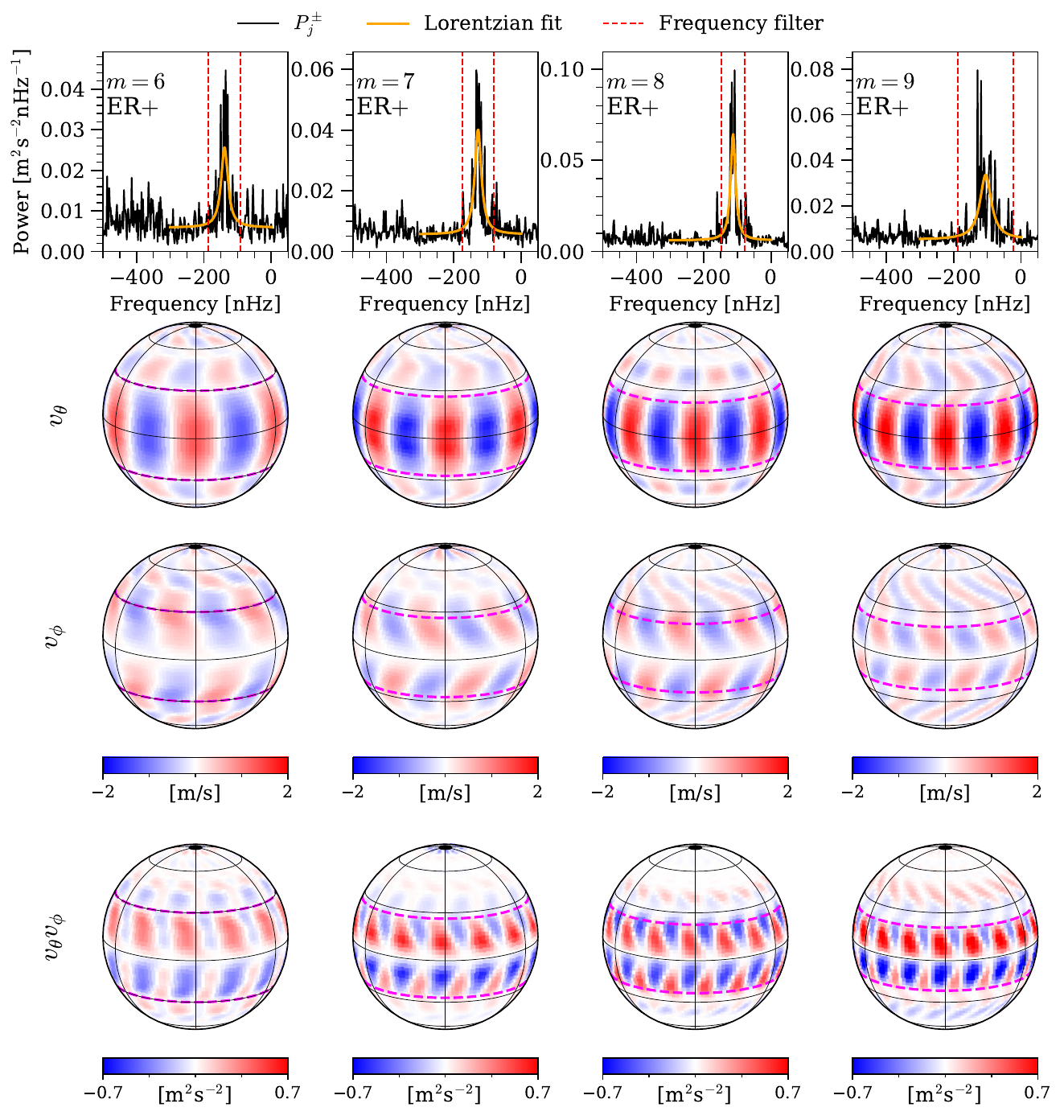}
\caption{Same as Fig.~\ref{f:ef1-4}  for the ER modes with $6\leq m\leq 9$.
}
\label{f:ef9-12}
\end{figure*}

\begin{figure*}
\sidecaption
\includegraphics[width=12cm]{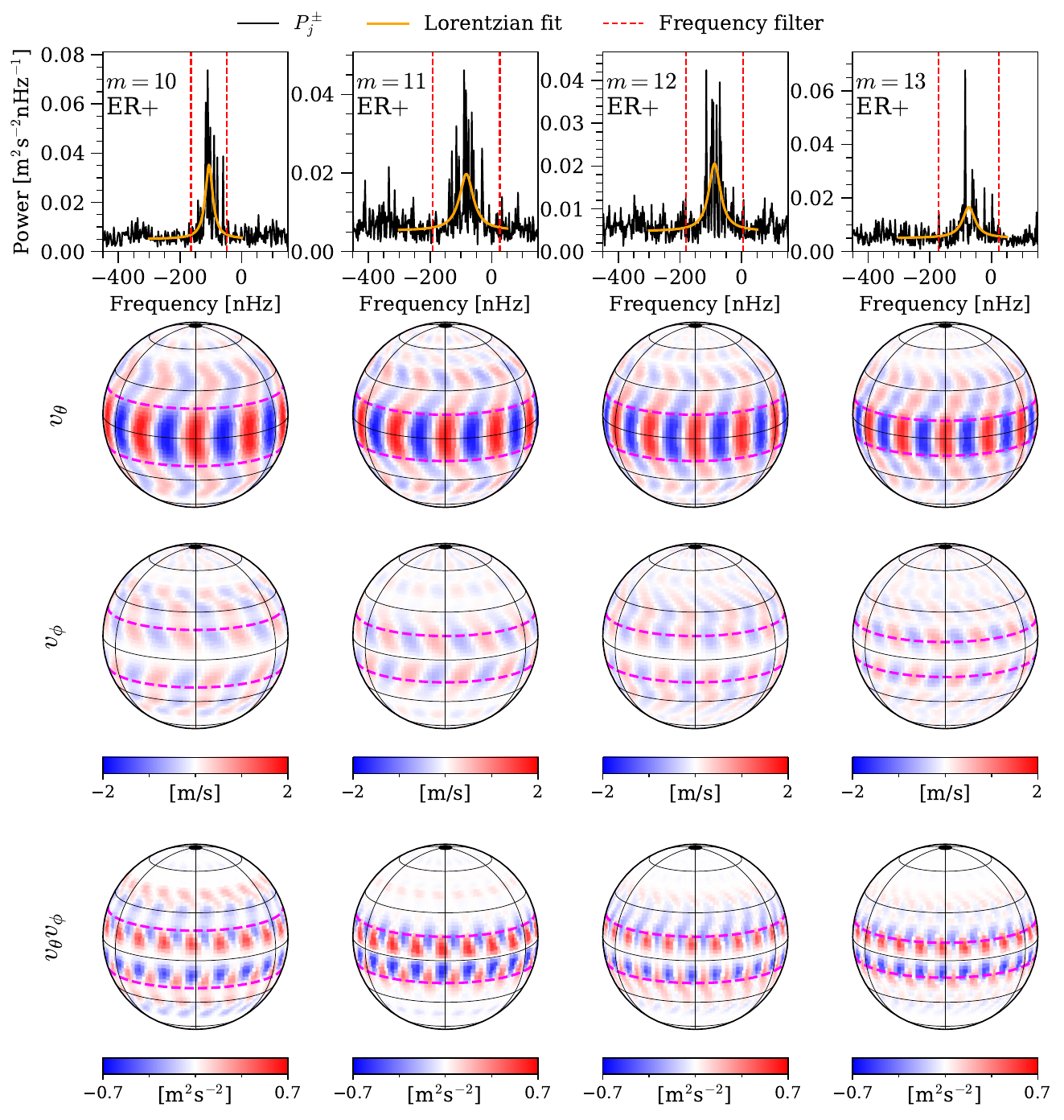}
\caption{Same as Fig.~\ref{f:ef1-4} for the ER modes with $10\leq m\leq 13$.
}
\label{f:ef13-16}
\end{figure*}

\begin{figure*}
\sidecaption
\includegraphics[width=12cm]{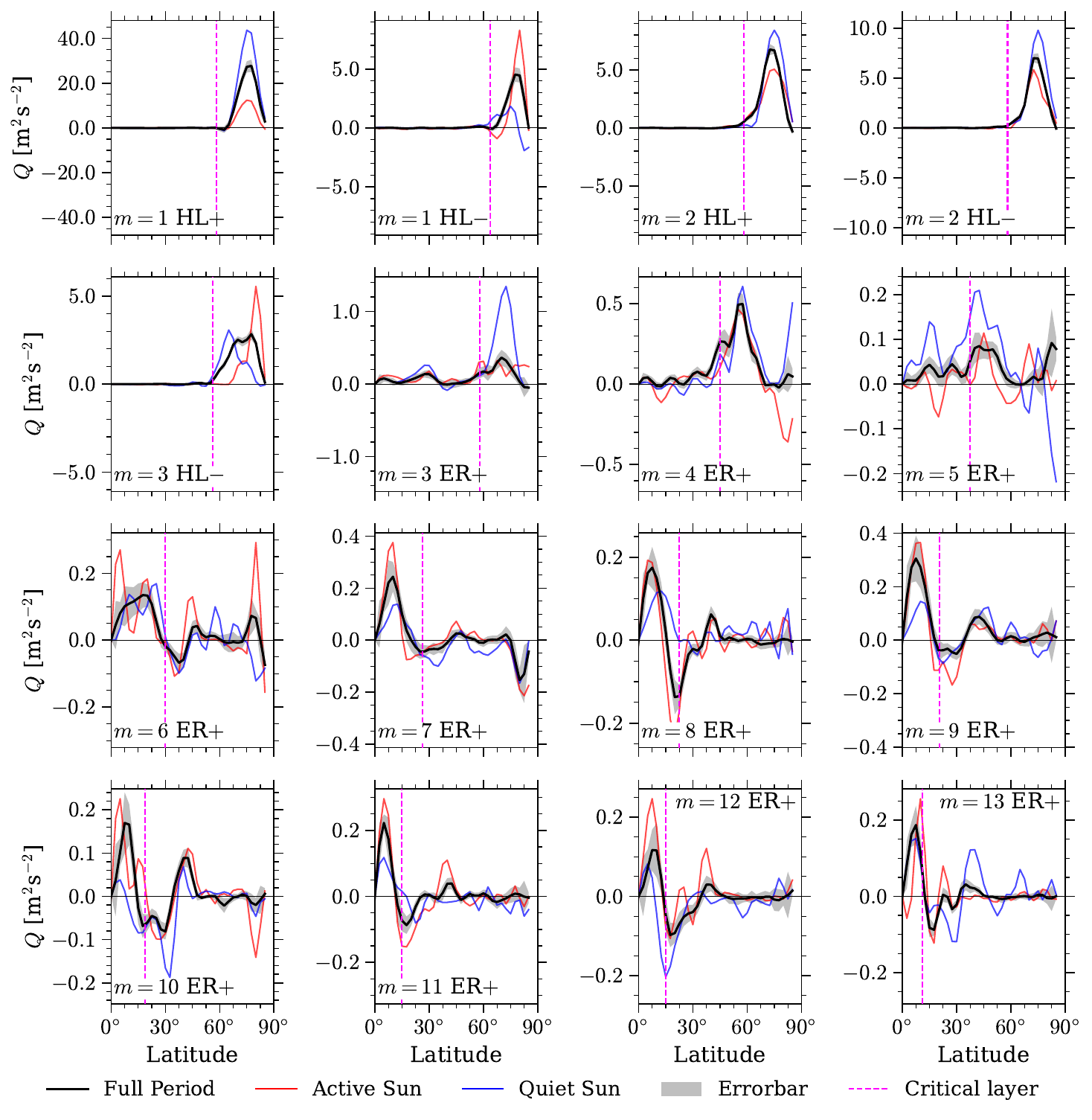}
\caption{
Profiles of the Reynolds stress, $Q(\theta)=\langle v_\theta v_\phi \rangle_\phi$, of the 16 selected inertial modes as functions of latitude.
Only the northern hemisphere is shown, as $Q$ is antisymmetric across the equator.
In each panel, black, red, and blue curves show the profiles computed using the full time period (2010--2025), near solar maximum (April 2021--April 2025), and near solar minimum (April 2016--April 2020).
The errors associated with black curves are denoted by gray shaded areas.
The latitudes of the critical layers, $\lambda_c$, are denoted by magenta lines.
}
\label{f:RS1D}
\end{figure*}

\begin{figure}
\centering
\includegraphics[width=0.8\hsize]{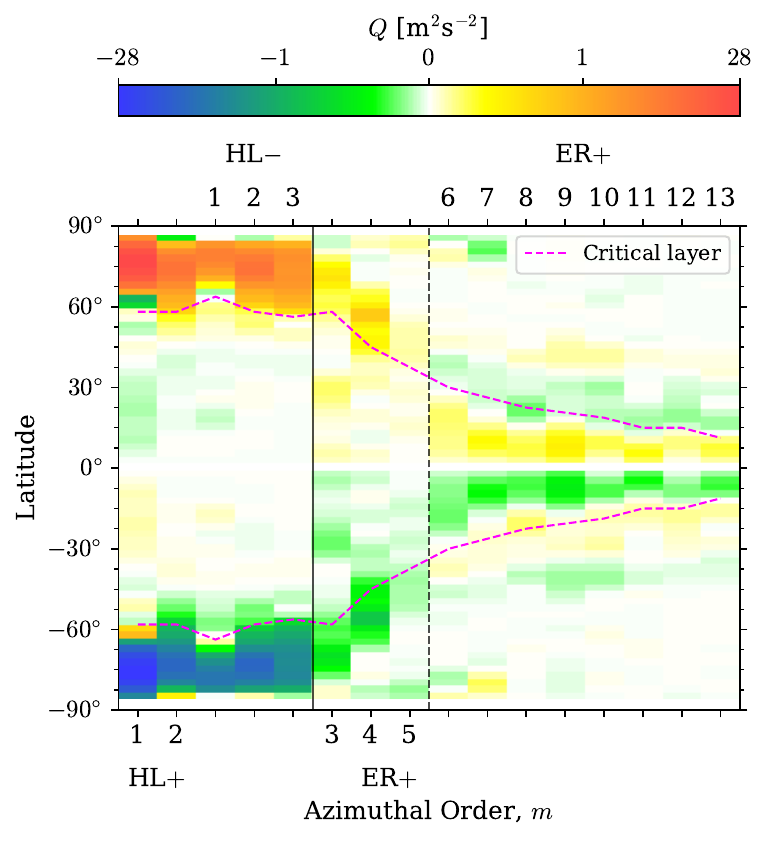}
\caption{Reynolds stress profiles of inertial modes extracted from full period 2010--2025.
The latitudinal Reynolds stress profiles for 16 inertial modes representing the same information as given by black curves in Fig.~\ref{f:RS1D}, along with their respective $\lambda_c$ (magenta).
The $m=1$ HL$+$ mode has an amplitude of $27.7$~m$^2$~s$^{-2}$.
The solid black line separates the HL modes on the left from ER modes on the right.
The dashed black line separates ER modes based on the qualitative differences between their Reynolds stress profiles.}
\label{f:rs_allto}
\end{figure}

\begin{figure}
\centering
\includegraphics[width=0.8\hsize]{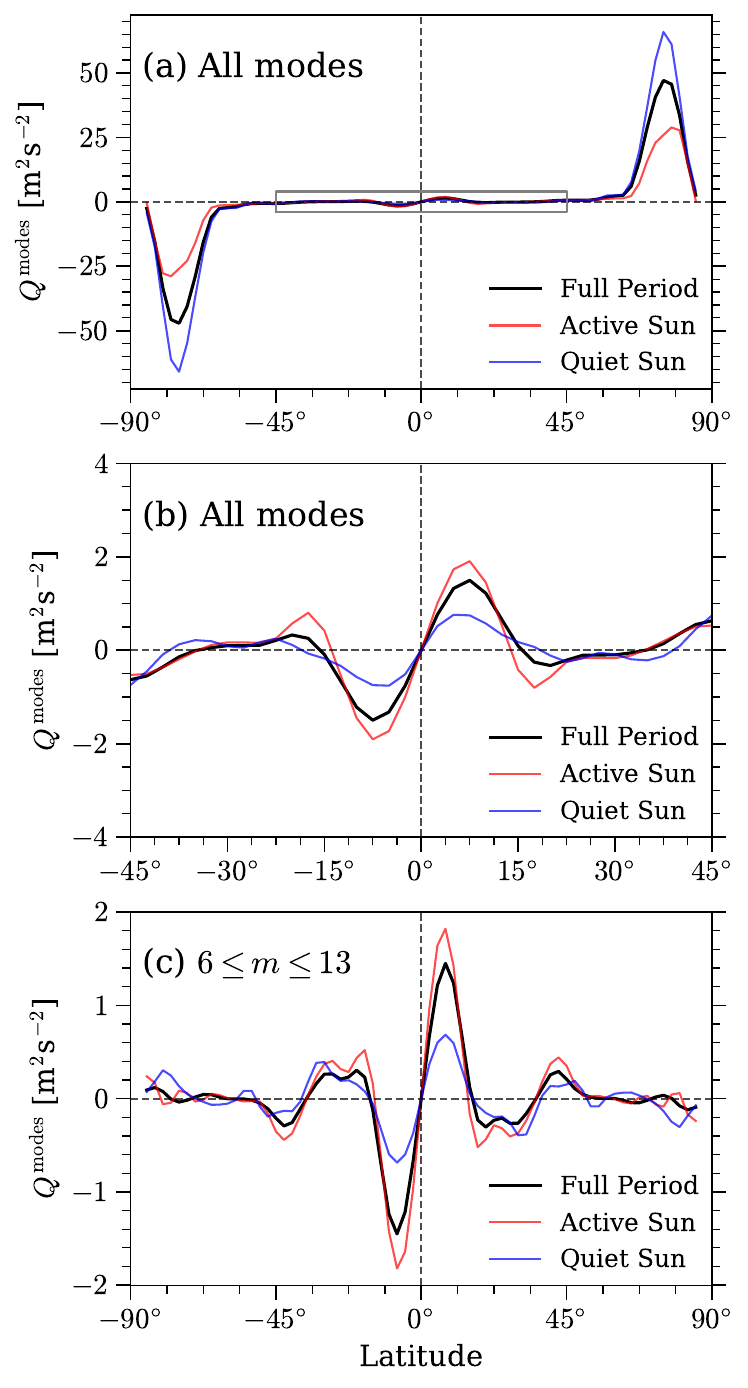}
\caption{
Sum of the Reynolds stresses of solar inertial modes, $Q^\mathrm{modes}$, defined in Eq.~\eqref{eq:RSmode}, for different phases of the solar cycle.
Black, red, and blue curves show the profiles computed using the whole period (2010--2025), near solar maximum (April 2021--April 2025), and near solar minimum (April 2016--April 2020), respectively.
(a) Full extent of the data showing the strong amplitudes of HL modes above $60\degree$ latitude.
Panel (b) shows the zoom-in focusing on the profiles at lower latitudes.
(c) Cumulative Reynolds stress of ER modes with $6\leq m\leq 13$. The high $m$ mode profiles show a change of sign near $\lambda_c$.
}
\label{f:RSsum}
\end{figure}

\section{Results and Discussion} \label{s:results}

Table~\ref{t:mode params} summarizes the basic parameters of the inertial modes analyzed in this study.
The mode frequencies, linewidths, and amplitudes are shown in Fig.~\ref{f:mode_parameters}. 
The mode parameters are known to vary with time \citep{waidele2023,liang2025,lekshmi2026}; therefore, some of our reported parameters may differ from previous studies \citep{liang2019,gizon2021}.
Figures~\ref{f:ef1-4}--\ref{f:ef13-16} show the observed velocity eigenfunctions of all the inertial modes analyzed in this study. 
In each figure, the first row shows the latitudinally averaged power spectra at fixed $m$ used to identify the modes, while the second and third rows show the eigenfunctions of the southward and prograde velocity components, $v_{\theta}$ and $v_{\phi}$. 
The bottom row shows their product, $v_{\theta}v_{\phi}$, which determines the Reynolds stress of each mode.

The latitudes of the critical layers $\lambda_c$ listed in Table~\ref{t:mode params} are the latitudes at which the modes corotate with the local differential rotation, i.e. $2\pi\nu_0 = m[\Omega(R_\odot, \lambda_c) - \Omega_{\rm Carr}]$, where $\nu_0$ is the mode frequency in the Carrington frame, $m$ is the azimuthal order, $\Omega$ is the sidereal angular velocity of the Sun, and $\Omega_{\rm Carr}/2\pi = 456$~nHz. 
For $\Omega$, we use the surface rotation rate from the two-dimensional RLS inversions provided by the JSOC data series \texttt{hmi.v\_sht\_2drls}, averaged over 2010--2025.

\subsection{High-latitude (HL) inertial modes} \label{s:HL}

The first five modes in Table~\ref{t:mode params} correspond to the HL modes. 
These modes exhibit their largest amplitudes at latitudes above $65^\circ$.
These HL modes have larger velocity amplitudes than those of the ER modes.
In particular, the $m=1$ HL$+$ mode has the largest amplitude ($11.6$~m~s$^{-1}$) and a narrow linewidth of $11.9$~nHz.
It is also noteworthy that, even for the same azimuthal order $m=1$, the antisymmetric HL$-$ mode has a much smaller amplitude ($5.6$~m~s$^{-1}$). 
This pronounced north-south asymmetry has also been reported in previous studies \citep[][]{gizon2021,liang2025,lekshmi2026}.

The surface velocity eigenfunctions of the HL modes display coherent large-scale structures near the poles.
The latitudes of the critical layers, $\lambda_c$, are marked by the dashed magenta lines in Fig.~\ref{f:ef1-4}. 
The power of the HL modes is concentrated at latitudes above $\lambda_c$.
In general, the latitudinal velocity eigenfunction $v_{\theta}$ is more confined toward the polar regions, whereas the longitudinal velocity $v_{\phi}$ has a broader latitudinal distribution that extends up to $\lambda_c$. 
It is also clear that the longitudinal velocity  $v_{\phi}$ eigenfunction displays a characteristic spiral pattern around the poles \citep[][]{hathaway2013,bogart2015,gizon2021,ding2025}. 
The spiral is oriented clockwise toward the pole when viewed from the north pole, and this behavior is found to be universal for all the HL modes.
We note that the orientation of these spiral patterns is closely related to the sign of the horizontal Reynolds stress discussed in Sect.~\ref{s:reynolds}.

For the most prominent $m=1$ mode, we find that the amplitude of $v_{\theta}$ shows an increasing trend up to $85\degree$. 
This trend likely reflects the presence of pole-crossing flows, as implied by previous observations \citep[][]{ulrich1993,liang2025} and numerical simulations \citep[][]{bekki2024}.

The $m=2$ and $m=3$ HL modes have velocity amplitudes of about $3$--$5.5$ m~s$^{-1}$, which are much weaker than that of the $m=1$ HL$+$ mode, but still greater than those of the equatorial Rossby modes.
For $m \ge 4$, the linewidths become significantly broader, resulting in poor signal quality. Therefore, we excluded these higher-$m$ HL modes from the present analysis.
As shown in Fig.~\ref{f:mode_parameters}a, the frequencies of the HL modes exhibit a nearly non-dispersive dependence on $m$. 
This behavior has led to the interpretation that these $m \ge 2$ signals do not represent independent eigenmodes, but rather harmonic components of the large-amplitude $m=1$ mode through nonlinear effects \citep[][]{mushtaq2026}.

\subsection{Equatorial Rossby (ER) modes} \label{s:ER}

Table~\ref{t:mode params} shows mode parameters corresponding to the ER modes at $m=3$--$13$. 
They have frequencies roughly following the classical dispersion relation of the sectoral Rossby modes \citep[][]{loeptien2018}.
The $m=3$ mode has the clearest signal with a linewidth of $10.5$~nHz. 
Most of them have velocity amplitudes of about $1$--$2$~m~s$^{-1}$, with linewidths generally increasing with $m$ (see also Fig.~\ref{f:mode_parameters}).

The surface velocity eigenfunctions of the ER modes are displayed in Figs.~\ref{f:ef5-8}--\ref{f:ef13-16}.
Their spatial eigenfunctions are characterized by a strong latitudinal velocity $v_{\theta}$ that is symmetric across the equator, whereas the antisymmetric longitudinal velocity $v_{\phi}$ is comparatively weak. 
As noted in previous studies~\citep[e.g.,][]{gizon2021}, the ER modes are quasi-toroidal.
This is supported by the fact that their horizontal flows involve a dominant radial vorticity that peaks at the equator, while exhibiting little or no significant horizontal divergence.

A closer inspection further reveals that the ER modes exhibit distinct properties in the ranges $3 \leq m \leq 5$ and $m \geq 6$.
For $3 \leq m \leq 5$, the velocity eigenfunctions of $v_\phi$ are comparable to $v_\theta$ in terms of amplitude. 
These eigenfunctions also show pronounced distortions at or above their $\lambda_c$, including spiral structures similar to that of the HL modes (Fig.~\ref{f:ef5-8}).
These distortions are more visible in the longitudinal velocity $v_{\phi}$.
As a consequence, the product $v_{\theta}v_{\phi}$ of these low-$m$ ER modes becomes positive (negative) in the northern (southern) hemisphere near or above $\lambda_c$.
In contrast, as shown in Figs.~\ref{f:ef9-12} and~\ref{f:ef13-16}, the ER modes with $m \geq 6$ do not exhibit the characteristic spiral distortions near their $\lambda_c$. 
This difference is particularly evident in $v_{\phi}$.
For such higher-$m$ ER modes, the eigenfunctions are largely confined to latitudes lower than $\lambda_c$, where the product $v_{\theta}v_{\phi}$ also tends to reach large values.
We also note that, on average, these higher-$m$ ER modes tend to have slightly larger mode amplitudes of $\sim 2$~m~s$^{-1}$ (Fig.~\ref{f:mode_parameters}c).

\subsection{Horizontal Reynolds stresses of inertial modes} \label{s:reynolds}

Figure~\ref{f:RS1D} presents the Reynolds stresses $Q$ (defined by Eq.~\eqref{eq:RS}) associated with all $16$ inertial modes examined in this study.
The latitudinal profiles of $Q$ are strictly antisymmetric, reflecting the enforced symmetry of the extracted mode eigenfunctions.
The high-latitude (HL) modes ($1 \leq m \leq 3$) display substantial Reynolds stresses at very high latitudes (typically $\gtrsim 60^\circ$). 
All HL modes exhibit positive (negative) Reynolds stresses, $Q > 0 \ (<0)$, in the northern (southern) hemisphere, implying an equatorward transport of angular momentum.
The sign of $Q$ is closely linked to the orientation of the spiral structures in the velocity eigenfunctions, which spiral clockwise toward the pole as viewed from the north pole.
The amplitudes of $Q$ attain their maxima at latitudes above $\lambda_c$.
In particular, the dominant $m=1$ HL$+$ mode attains a very large Reynolds stress amplitude of $Q = 44~\mathrm{m}^2~\mathrm{s}^{-2}$ during solar minimum.

The equatorial Rossby (ER) modes exhibit weaker Reynolds stresses with latitudinal profiles that are more complex than those of the HL modes.
As outlined in Sect.~\ref{s:ER}, the ER modes can be subdivided into two distinct groups.
For $3 \leq m \leq 5$, the associated Reynolds stresses peak near or slightly above $\lambda_c$.
Analogous to the HL modes, $Q$ tends to be positive (negative) in the northern (southern) hemisphere, again corresponding to an equatorward angular-momentum flux.
For $m \geq 6$, by contrast, the Reynolds stresses tend to peak at latitudes below $\lambda_c$ and exhibit pronounced sign reversals in the vicinity of $\lambda_c$. 
In the northern hemisphere, $Q$ is positive at latitudes below $\lambda_c$ and becomes negative at latitudes above $\lambda_c$.
These characteristics are clearly illustrated in Fig.~\ref{f:rs_allto}, which shows the Reynolds stresses of all modes in a side-by-side comparison.

As reported in previous studies \citep[][]{waidele2023,liang2025}, both the HL and ER mode families exhibit pronounced solar-cycle variability. 
To characterize the cycle dependence of the associated Reynolds stresses, we separately analyze two four-year observational intervals: April 2021--April 2025, corresponding to near-maximum solar activity, and April 2016--April 2020, corresponding to near-minimum activity.
For each interval, we extract the inertial modes and compute the Reynolds stresses using the methodology described above. 
The resulting Reynolds-stress profiles for the active and quiet phases are displayed in Fig.~\ref{f:RS1D} as red and blue curves, respectively.
The $m=1$ HL$+$ mode, the two $m=2$ HL$\pm$ and the $m=3$ ER$+$ mode are stronger close to the solar minimum.
The $m=1$ HL$-$ mode, $m=3$ HL$-$ mode, and most ER modes with $m\geq6$ are stronger close to the solar maximum.
These findings are consistent with results of \citet{lekshmi2026} showing that $m=1$ HL$+$ mode amplitude is anticorrelated with sunspot number, whereas ER-mode amplitudes exhibit a positive correlation with sunspot number.

To further highlight the distinctions among the different mode families, we compare their Reynolds-stress profiles side by side in Fig.~\ref{f:rs_allto}, together with the corresponding latitudes of their critical layers $\lambda_c$. This comparison suggests three distinct groups:
(i) HL modes (both symmetric and antisymmetric) have $Q$ concentrated predominantly at latitudes poleward of their respective $\lambda_c$, with negligible Reynolds stress equatorward of $\lambda_c$.
(ii) ER modes with $3\leq m\leq 5$ have $Q$ concentrated primarily above and in the vicinity of $\lambda_c$, while maintaining a weaker but non-negligible signal below $\lambda_c$.
(iii) ER modes with $6\leq m\leq 13$ have $Q$ concentrated mainly between the latitudes of their critical layers in the two hemispheres; crossing $\lambda_c$ toward higher latitudes is accompanied by a sign reversal in $Q$, as expected when Rossby modes are modified by nonlinear interactions \citep{vallis}.

Figure~\ref{f:RSsum}a and b present the latitudinal profile of the total Reynolds stress, $Q^\mathrm{modes}$, obtained by summing the contributions from all $16$ identified inertial modes:
\begin{equation}
Q^\mathrm{modes} = \sum_{\rm modes} Q. \label{eq:RSmode}
\end{equation}
Although only a subset of the full inertial-mode spectrum is considered, $Q^\mathrm{modes}$ in the near-surface layers of the Sun is not expected to change substantially when the remaining inertial modes are taken into account.
The quantity $Q^\mathrm{modes}$ is largely dominated by the high-latitude (HL) modes with large amplitudes, in particular the $m=1$ HL$+$ mode.
Consequently, the magnitude of $Q^\mathrm{modes}$ at high latitudes ($\gtrsim 60^\circ$), where the HL modes prevail, exceeds that at low latitudes ($\lesssim 30^\circ$, Fig.~\ref{f:RSsum}b), where equatorial Rossby (ER) modes contribute, by more than an order of magnitude.
Furthermore, the relative contributions at low and high latitudes exhibit a pronounced dependence on the solar cycle: during epochs of elevated magnetic activity the amplitude increases at low latitudes and decreases at high latitudes, while the opposite trend is observed during magnetically quiet phases.

The Reynolds stresses $Q^\mathrm{modes}$ are similar in both spatial structure and amplitude to those reported by \citet{hathaway2021}, who employed flow maps derived from local correlation tracking of supergranules in HMI Dopplergrams. They reported a predominantly antisymmetric Reynolds-stress profile with amplitudes up to $\sim 50$~m$^2$~s$^{-2}$, indicative of an equatorward transport of angular momentum.

In Fig.~\ref{f:RSsum}c, we isolate $Q^\mathrm{modes}$ for ER modes with azimuthal orders $6\leq m\leq 13$. Their amplitude is correlated with solar activity, increasing from $\sim 1$~m$^2$~s$^{-2}$ during the quiet-Sun interval to $\sim 2$~m$^2$~s$^{-2}$ during the active phase. However, the location of this maximum stays close to $8\degree$ latitude.
These modes exhibit a distinct sign reversal in $Q$ near $15\degree$ latitude, followed by a second sign change near $37\degree$; the resulting latitudinal profile is reminiscent of an exponentially decaying wave.
The measured horizontal Reynolds-stress profiles of the ER modes differ in several important respects from the theoretical predictions of \citet{gizon2020b}. In particular, for $m\geq 6$ the observed profiles are more narrowly confined in latitude, undergo a sign change at $\lambda_c$ and a second reversal at higher latitudes, and possess significantly smaller amplitudes. To assess whether these mode stresses matter dynamically, we compare them against the total surface Reynolds stress.

\subsection{Comparison with other sources of Reynolds stresses} \label{s:TO}

\begin{figure*}
\sidecaption
\includegraphics[width=12cm]{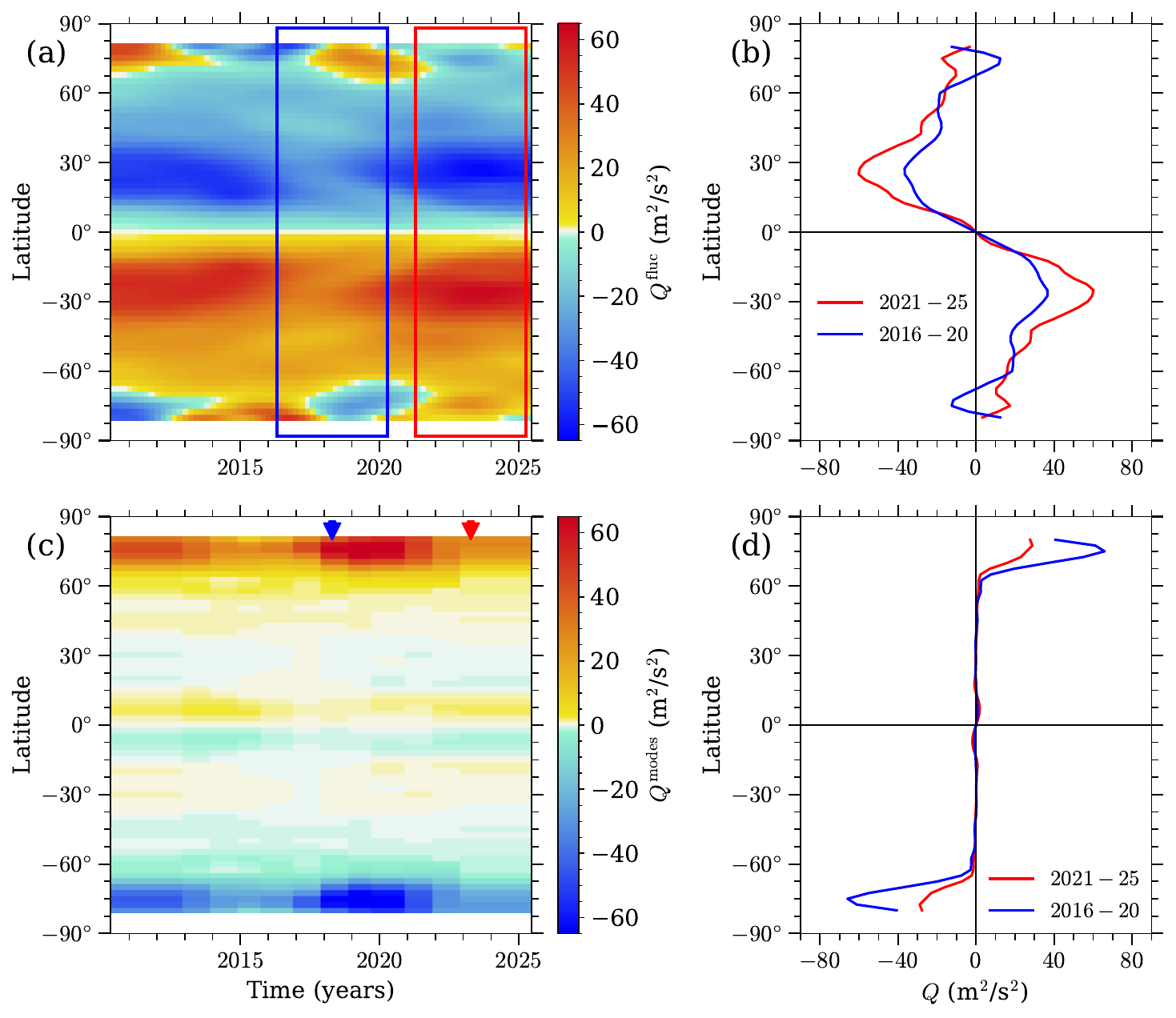}
\caption{
Temporal evolution of the Reynolds stress.
(a) Reynolds stress $Q^\mathrm{fluc}$ computed from the unfiltered horizontal velocity flow maps at a depth of $0.997~R_\odot$. The component anti-symmetric across the equator is shown.
(b) Latitudinal profiles of $Q^\mathrm{fluc}$ averaged over two 4-year windows: April 2021--April 2025 (red) and April 2016--April 2020 (blue).
(c) Cumulative Reynolds stress $Q^\mathrm{modes}$ from the 16 selected inertial modes.
(d) Latitudinal profiles of $Q^\mathrm{modes}$ averaged over two 4-year windows: April 2021--April 2025 (red) and April 2016--April 2020 (blue).
}
\label{f:RS_totmodes}
\end{figure*}

\begin{figure}
\centering
\includegraphics[width=0.8\hsize]{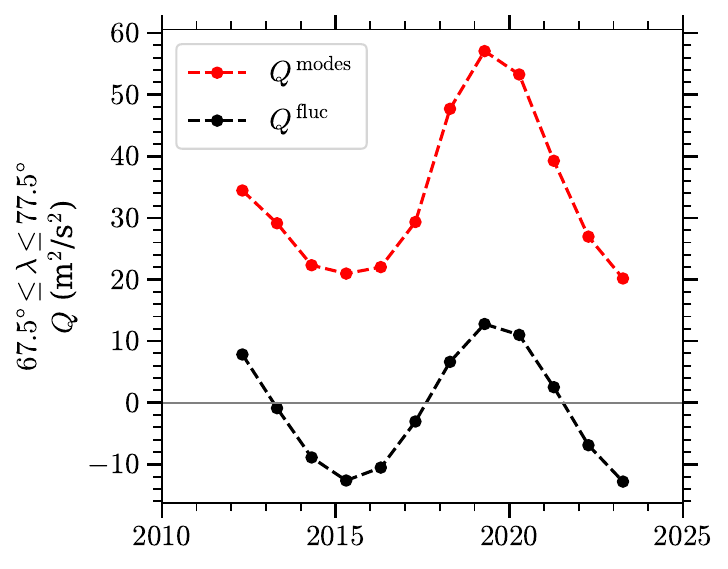}
\caption{$Q^\mathrm{fluc}$ and $Q^\mathrm{modes}$ averaged over latitudes $67.5\degree \leq \lambda \leq 77.5\degree$, with $Q^\mathrm{fluc}$ averaged over the same 4-year chunks used for $Q^\mathrm{modes}$.
The two are strongly correlated and peak at solar minimum, when the HL modes are strongest.}
\label{f:RStime}
\end{figure}

\begin{figure}
\centering
\includegraphics[width=0.8\hsize]{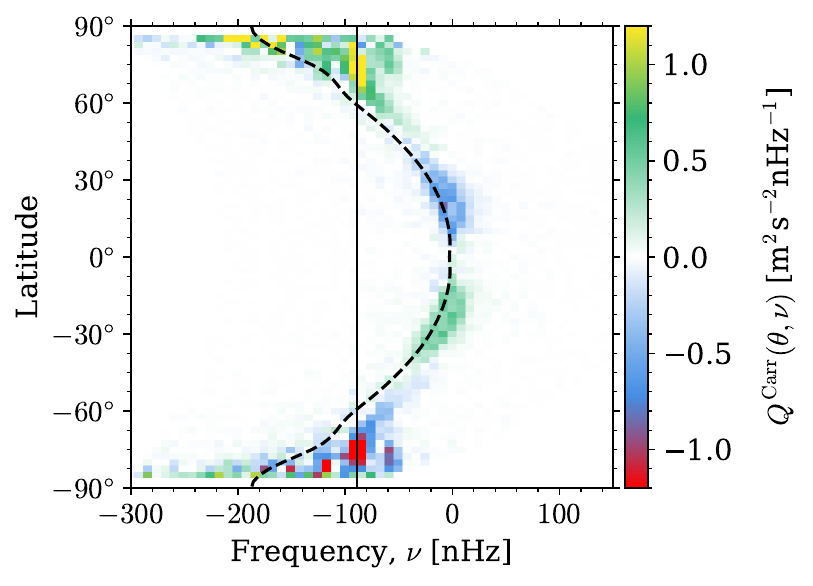}
\caption{Horizontal Reynolds stress in Carrington frame $Q^\mathrm{Carr}$. 
The procedure to obtain this is described by Figs.~\ref{f:qmask} and \ref{f:rs_unfiltcarr}.
This plot is close to being antisymmetric across the equator.
The strong negative (positive) $Q$ in the southern (northern) hemisphere comes from HL modes, mostly the $m=1$ HL$+$ mode.
At low-latitudes, the inertial modes are insignificant, and we see active region related flows dominating.}
\label{f:RS_ucf}
\end{figure}

In this section, we discuss the extent to which the time-varying Reynolds stresses of inertial modes may be significant in convection zone dynamics. One of the direct consequences of Reynolds stresses can be seen in the equation of angular momentum balance, which describes the temporal evolution of differential rotation at the solar surface \citep[e.g.,][]{ruediger89, miesch2005}
\begin{equation}\label{eq:amb_simple}
r\sin\theta \frac{\partial \Omega}{\partial t} = - \frac{1}{r \sin^2{\theta}} \frac{\partial}{\partial \theta}\left[\sin^2\theta ( Q^\mathrm{mean} + Q^\mathrm{fluc}) \right] + [..],
\end{equation}
where $[..]$ represents the contribution of magnetic effects and turbulent diffusivity, as well as the radial derivatives, which are not considered here, and 
\begin{equation}
Q^\mathrm{mean} = u_\mathrm{mer} u_{\rm zonal},
\end{equation}
is the advection of angular momentum by the axisymmetric ($m=0$) flows, in this case, the interaction between the meridional circulation ($u_\mathrm{mer}$) and the zonal flows ($u_{\rm zonal}$). 
The component of Eq.~\eqref{eq:amb_simple}, that we want to focus on is $Q^\mathrm{fluc}$, the Reynolds stress contribution of other flows such as active region inflows, turbulent convection, and inertial modes ($Q^\mathrm{modes}$) as seen on the surface of the Sun through ring-diagrams.
We evaluate
\begin{equation}
Q^\mathrm{fluc}(\theta, t) = \langle u'_\theta(\theta, \phi, t) u'_\phi(\theta, \phi, t) \rangle_\phi,
\label{eq:qfluc}
\end{equation}
where $Q^\mathrm{fluc}$ is the longitudinally averaged Reynolds stress from horizontal flows $u'_\theta$ and $u'_\phi$ (from Sect.~\ref{symm}), and we show its antisymmetric component (across the equator) in Fig.~\ref{f:RS_totmodes}a. 
Both quantities $Q^\mathrm{fluc}$ and $Q^\mathrm{mean}$ undergo a similar smoothing procedure. They are binned in time to a cadence of $\sim72$~days, and then smoothed in both latitude and time using a Gaussian filter with $\sigma=(360~{\rm days},\,2.5\degree)$, truncated at $4\sigma$ in each dimension.

To investigate the temporal variations of the inertial modes and how much they contribute to $Q^\mathrm{fluc}$, we perform a chunk analysis. We divide the flow maps into 4-year chunks with the centroid of successive chunks shifted by one year, and apply the procedure described in Sect.~\ref{s:flw} and \ref{s:fee} to each chunk. This provides us with the horizontal velocity eigenfunctions $\mathbf{v}(\theta, \phi; t)$ for each mode corresponding to a time chunk $t$.
Figure~\ref{f:RS_totmodes}c shows the sum of the Reynolds stress from all the $16$ inertial modes, $Q^\mathrm{modes}$, as a function of time and latitude, and it is defined as
\begin{equation}
Q^\mathrm{modes}(\theta; t) = \sum_{\rm modes} \langle v_{\theta}(\theta, \phi; t) v_{\phi}(\theta, \phi; t) \rangle_\phi.
\end{equation}
As discussed in Sect.~\ref{s:reynolds}, $Q^\mathrm{modes}$ exhibits distinct temporal evolution patterns at low and high latitudes.

Figure~\ref{f:RS_totmodes}a shows that $Q^\mathrm{fluc}$ is mostly negative (positive) in the northern (southern) hemisphere.
There are small pockets where the signs change and comparing to Fig.~\ref{f:RS_totmodes}c, we can see that it coincides with the increase of power of inertial modes, specifically the HL modes, which explains why these pockets are only seen at latitudes above $60\degree$. This implies that the HL inertial modes have visible impact at these latitudes.
At lower latitudes, however, $Q^\mathrm{modes}$ is dominated by ER modes (see also Fig.~\ref{f:rs_allto}). However, it is completely overshadowed by other flows, as these signals do not turn up in Fig.~\ref{f:RS_totmodes}a.

We also show latitudinal cuts corresponding to different time periods for quantities $Q^\mathrm{fluc}$ and $Q^\mathrm{modes}$ in Figs.~\ref{f:RS_totmodes}b and d, respectively.
In both these figures, the blue curve corresponds to a four-year time period, April 2016--April 2020, of low solar activity. On the other hand, the red curve corresponds to April 2021--April 2025, a period of comparatively higher solar activity.
In Fig.~\ref{f:RS_totmodes}b, the blue and red curves are the time averaged $Q^\mathrm{fluc}$ in their respective durations mentioned above.
The blue curve, with a peak amplitude of $40$~m$^2$s$^{-2}$, shows the sign flip of Reynolds stress that can be directly attributed to HL modes (see also Fig.~\ref{f:RStime}). 
The $Q^\mathrm{fluc}$ averaged for this is highlighted in blue box on Fig.~\ref{f:RS_totmodes}a, where we look in the time period least affected by activity. It is negative in northern hemisphere up to $65\degree$ latitude.
This shows that turbulent convection, at the surface, transports angular momentum poleward.
The red curve here has much stronger stresses between 10$\degree$--35$\degree$ latitudes, due to stronger active region inflows, while the contribution of HL modes is smaller. It has a peak amplitude of $60$~m$^2$s$^{-2}$.
Between latitudes $35\degree$ and $65\degree$, there is a minor difference between the red and blue curves, which shows that the turbulent convection dominating here is largely unchanged in the observed time periods.
Both active (red) and quiet (blue) period $Q^\mathrm{fluc}$ peak near $25\degree$ latitude.

Our results of $Q^\mathrm{fluc}$ have some similarities with those of \citet{hanasoge2016}. The Reynolds stresses they show are positive (negative) in southern (northern) hemisphere, but with a much larger amplitude ($\sim 300$~m$^2$s$^{-2}$).
Lower amplitudes from ring-diagrams can be attributed to the larger tiles over which our flows are averaged.

In Fig.~\ref{f:RS_totmodes}b and d, we see certain correlations between $Q^\mathrm{fluc}$ and $Q^\mathrm{modes}$, at latitudes above $60\degree$. The red curve is the total Reynolds stress of the modes from extracted eigenfunctions for the period from April 2021 to April 2025, when the Sun is more active compared to the blue curve extracted from flow maps between April 2016 and April 2020. Here, we can see that the amplitude of red curve is smaller at $30$~m$^2$s$^{-2}$ compared to blue at $70$~m$^2$s$^{-2}$. Both the red and blue curves of Fig.~\ref{f:RS_totmodes}d peak near $75\degree$ latitude.

Figure~\ref{f:RStime} compares $Q^\mathrm{fluc}$ and $Q^\mathrm{modes}$, where $Q^\mathrm{fluc}$ is averaged over the multiple four-year chunks shifted by one year, same time period for which the inertial modes are extracted for $Q^\mathrm{modes}$. Here, we can see a direct correlation in their behavior at high latitudes. This shows that the changes in $Q^\mathrm{fluc}$ can be attributed directly to changes in $Q^\mathrm{modes}$, specifically the HL modes.

\subsection{Reynolds stresses in frequency space}

Figure~\ref{f:RS_ucf} allows us to analyze the different contributions to the Reynolds stress in frequency space, without isolating individual inertial modes (see also Appendix~\ref{sec:RS_unfilt}). The frequency axis is rescaled so that the latitudes of the critical layers of all modes coincide, allowing us to examine the exchange of angular momentum between the non-axisymmetric flows and the mean flows. 
Distinct regions of concentrated Reynolds stress are visible. Because the frequency axis is rescaled, all inertial modes are shifted from $\nu_0$ to $\nu_0/m$. However, since the total Reynolds stress is dominated by the $m=1$ HL$+$ mode at high latitudes and by active region flows at low latitudes, the other inertial modes are difficult to discern. 
The Reynolds stress associated with the HL modes and that associated with active-region flows have opposite signs in the same hemisphere, and only the former transports angular momentum equatorward. The figure thus illustrates how different mechanisms govern the dynamics in different regions of the Sun.

The Reynolds stress associated with turbulent convection is present at all latitudes, is mostly time independent, and transports angular momentum poleward. The Reynolds stress associated with active-region inflows is correlated with solar activity, confined to latitudes below $30\degree$, and also transports angular momentum poleward. In contrast, the Reynolds stress associated with the inertial modes is dominated by the HL modes, and it transports angular momentum equatorward.

\section{Conclusions} \label{s:discussion}

We present horizontal velocity eigenfunctions and mode parameters for 16 selected inertial modes, comprising 5 HL modes and 11 ER modes. The mode parameters are known to vary with time \citep{lekshmi2026}. Hence, the values reported here may be different from those reported in other studies due to the longer time period used. Using these eigenfunctions, we compute horizontal Reynolds stresses on the solar surface.

All HL modes exhibit Reynolds stresses $Q=\langle v_\theta v_\phi\rangle_\phi$ that are positive (negative) in the northern (southern) hemisphere, corresponding to equatorward angular momentum transport. 
These Reynolds stresses are concentrated above the latitudes of their critical layers $\lambda_c$ located near $60\degree$. Among them, the $m=1$ HL$+$ mode is by far the dominant contributor to the cumulative Reynolds stress $Q^\mathrm{modes}$.
The high-latitude Reynolds stresses ($30$--$70$~m$^2$~s$^{-2}$) associated with the HL modes exhibit a clear solar-cycle dependence, being stronger during the solar minimum and weaker during the solar maximum. The amplitude of this variation is sufficiently large to account for the observed temporal variation of the total Reynolds stress $Q^\mathrm{fluc}$, including the contribution from the background turbulent convection.

The horizontal Reynolds stress analysis reveals that the ER modes can be divided into two distinct categories depending on $m$. 
Modes with $3\leq m\leq 5$ exhibit Reynolds stress profiles similar to those of the HL modes, with Reynolds stresses transporting angular momentum equatorward and peaking above $\lambda_c$. 
This behavior arises because, above $\lambda_c$, the low-$m$ ER modes exhibit spiral flow structures similar to those of the HL modes.

However, as $m$ increases and $\lambda_c$ shifts to $30\degree$ and below, the Reynolds stress profiles of the modes start peaking below their $\lambda_c$.
For a mode with $m\geq6$, the Reynolds stress changes sign twice in a hemisphere, once near the latitude of the critical layer, and once again at higher latitudes: the Reynolds stress transports angular momentum equatorward below $\lambda_c$, poleward just above $\lambda_c$, and equatorward again at even higher latitudes, with the signal weakening as latitude increases. 
This sign reversal is a new feature not present in previous linear calculations~\citep[e.g.,][]{gizon2020b}.

At low latitudes, the cumulative Reynolds stress $Q^\mathrm{modes}$ is dominated by the contribution from the $m \geq 6$ ER modes, and exhibits a solar-cycle variation opposite to that at high latitudes, being weaker during the solar minimum and stronger during the solar maximum.
Its amplitude ($1$--$2$~m$^2$~s$^{-2}$) is much smaller than that of the total Reynolds stress $Q^\mathrm{fluc}$ ($40$--$60$~m$^2$~s$^{-2}$), which is dominated at these latitudes by convection and active-region inflows and mainly transports angular momentum poleward.
Our results thus suggest that the dynamical contribution of the ER modes to the angular momentum transport is limited, at least at the solar surface.
In contrast, we have shown that HL modes do play an important role in transporting angular momentum at the surface.

\begin{acknowledgements}
Author contributions: Y.~M. and L.~G. designed research. 
R.~B. computed the full set of  RDA 5$^\circ$-tile flow maps. Y.~M. carried out the analysis of the data and produced all the figures. Z.-C.~L. contributed to  the analysis of the  power spectra. Y.M. wrote the initial draft and all authors contributed to the final manuscript.  
We thank R.~Cameron, D.~Fournier, N.~Joshi, and J.~Schou for helpful discussions and comments. 
The HMI data are courtesy of NASA/SDO and the HMI Science Team. 
Y.~M. is a member of the International Max Planck Research School for Solar System Science at the University of Göttingen.
We acknowledge partial financial support from the ERC Synergy Grant WHOLESUN (grant agreement No. 810218) to L.~G. 
\end{acknowledgements}

\bibliographystyle{aa}
\bibliography{ref_list}

\appendix

\onecolumn
\section{Eigenfunction smoothing and estimation of random errors}\label{s:err}

\begin{figure*}[h]
\centering
\includegraphics[width=0.9\textwidth]{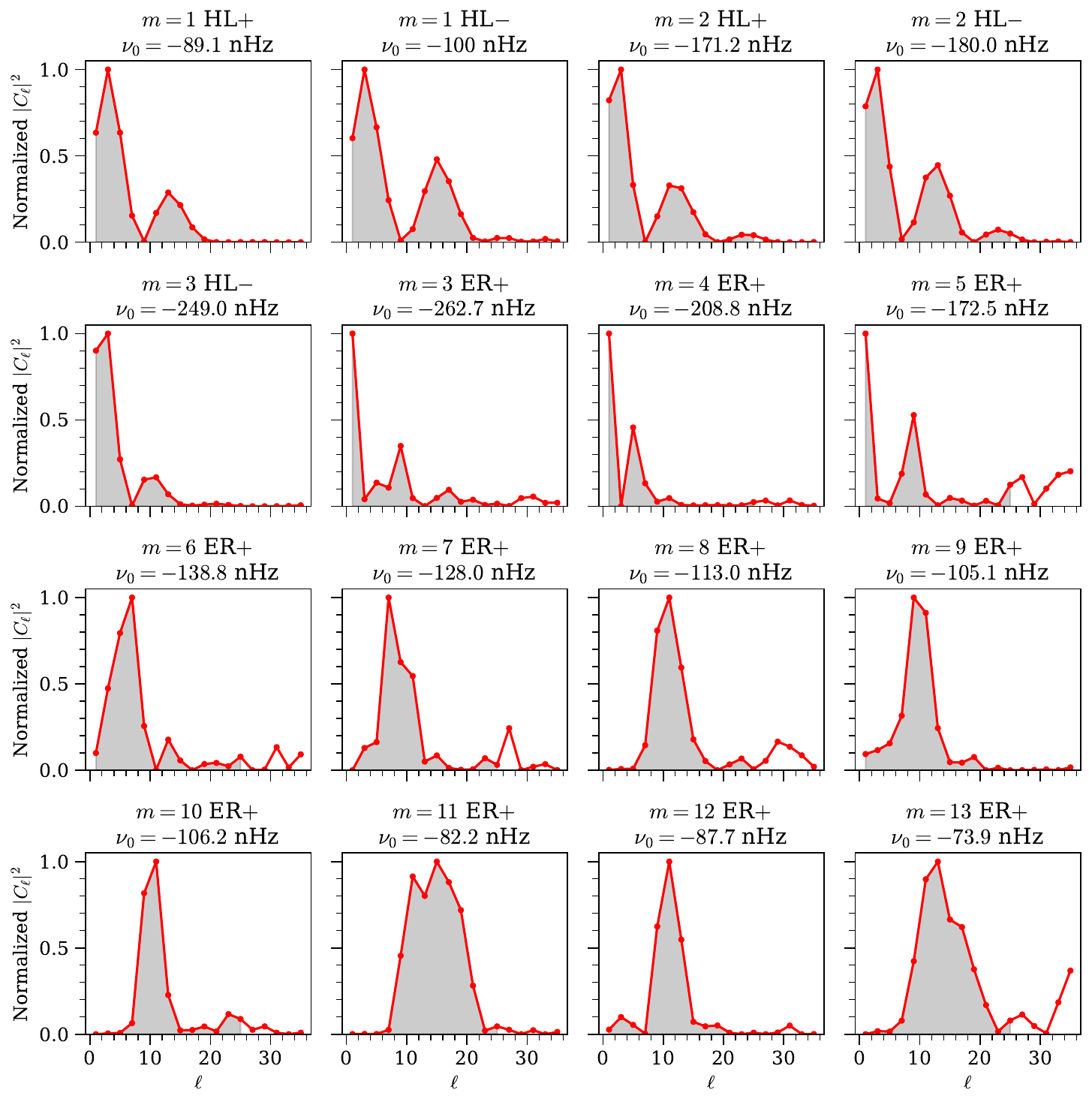}
\caption{Normalized power $|C_\ell|^2$ of the Legendre coefficients of the horizontal Reynolds stress profile $Q(\theta)$, for each of the 16 inertial modes. The power of each mode is normalized to its maximum. The shaded region marks the coefficients retained below the cutoff, $\ell \le \ell_c = 25$. Panel titles give the azimuthal order, the mode type (`HL' for high-latitude and `ER' for equatorial Rossby modes, with `+' and `$-$' denoting the north-south symmetry of $\zeta_r$), and the fitted frequency $\nu_0$.}
\label{f:legendre_coeffs}
\end{figure*}

Inertial modes are global-scale waves, so their eigenfunctions should vary smoothly with latitude. Small-scale fluctuations come from convection and noise. To remove them, we expand each velocity eigenfunction, $v_\theta(\theta)$ and $v_\phi(\theta)$, in Legendre polynomials and keep only degrees up to a cutoff $\ell_c$.

The velocity eigenfunctions are too noisy to show clearly where their signal ends. We therefore use the Reynolds stress profiles $Q(\theta)$, which are smoother. Figure~\ref{f:legendre_coeffs} shows the normalized power $|C_\ell|^2$ of the Legendre coefficients of $Q(\theta)$ for all 16 modes. In most cases, the power vanishes above $\ell \approx 25$. We adopt $\ell_c = 25$ for all modes and apply it to both velocity components.

We estimate the uncertainty from the velocity coefficients above $\ell_c$. We replace each of these coefficients with a random value: its amplitude follows a $\chi^2$ distribution with mean equal to the original amplitude, and its phase is uniform over $[0, 2\pi)$. The inverse Legendre transform then gives one realization of the eigenfunction, from which we compute $Q$. We repeat this 1000 times. The standard deviation of $Q$ across realizations, at each latitude, is the reported uncertainty.

\twocolumn
\section{Correcting for Incomplete Coverage}
\label{sec:ac}

\begin{figure}[h]
\centering
\includegraphics[width=0.8\linewidth]{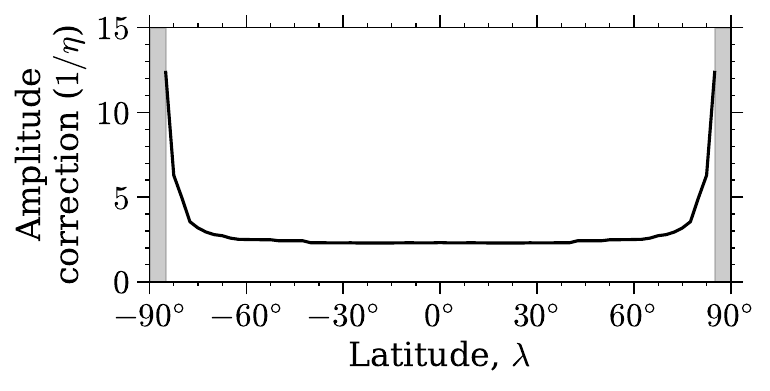}
\vspace{-0.2cm}
\caption{Correction factor $\eta^{-1}$ used in Eq.~\eqref{eq:FT2D}, where $\eta$ is the fraction of available data at each latitude.
It varies with latitude because the Sun's rotation axis is inclined with respect to the ecliptic.
There are no data in the gray region.
}
\label{f:ac}
\end{figure}

\section{Reynolds stresses in Carrington frame}\label{sec:RS_unfilt}

\begin{figure}[h]
\centering
\includegraphics[width=\hsize]{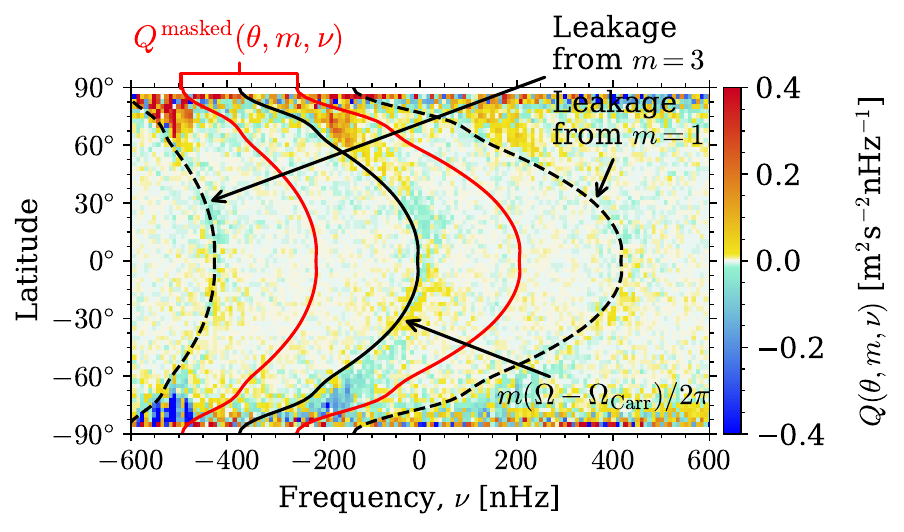}
\vspace{-0.5cm}
\caption{Filter used for computation of $Q^\mathrm{Carr}$.
The black curve shows the rotation curve for the azimuthal order under consideration, which in this example is $m=2$.
The dashed black curves show the window function leakage of differential rotation from neighboring $m$.
The red curves are the midpoint of the black curve and the leakages, and the region between these two red curves is retained for further computation of $Q^\mathrm{Carr}$.}
\label{f:qmask}
\end{figure}

\begin{figure}
\centering
\includegraphics[width=0.9\hsize]{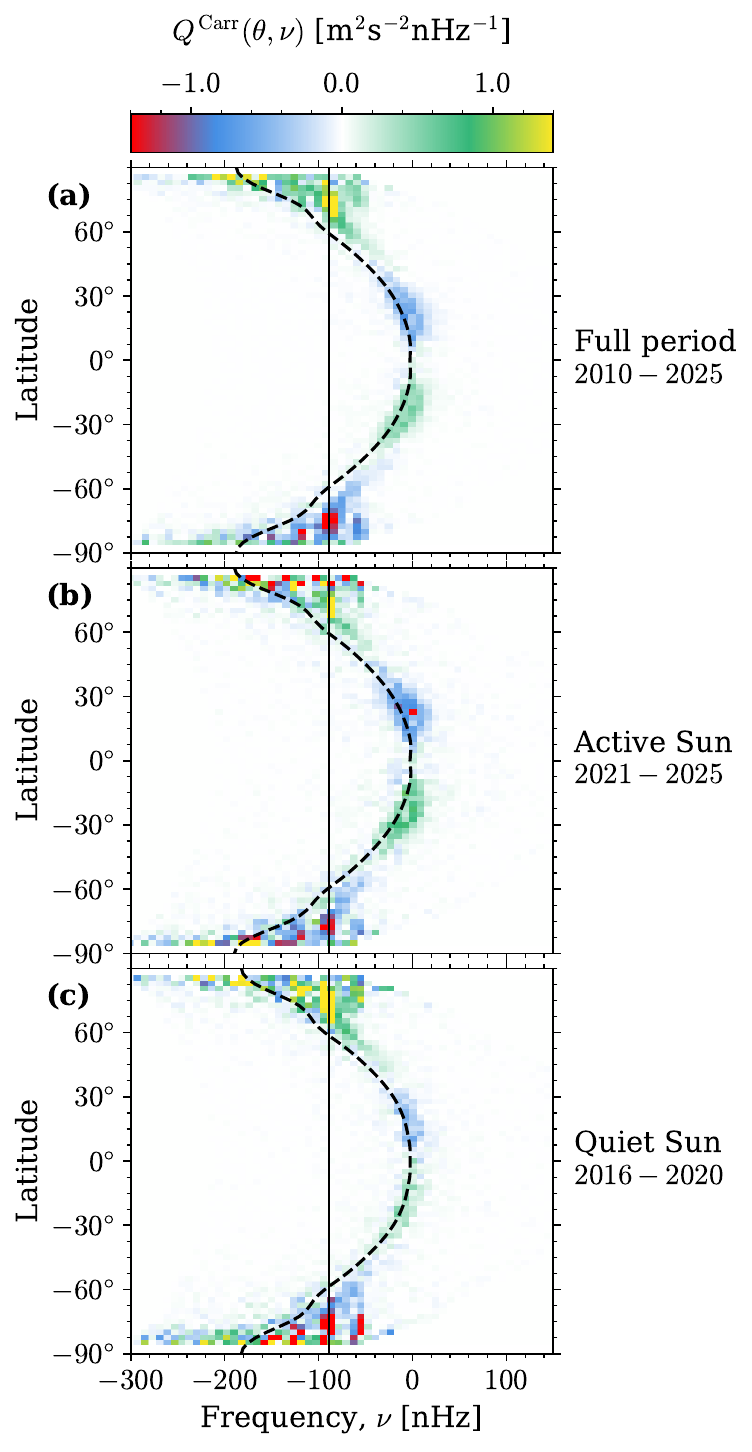}
\caption{Reynolds stress in Carrington frame. It is obtained from the 2D Fourier transform of horizontal velocity components (see Eq.~\eqref{eq:FT2D}), and summed over $m$, over three time periods full period (a), solar maximum (b), and solar minimum (c).
The black dashed line shows the differential rotation.
The solid black line highlights the location of $m=1$ HL$+$ mode at $-89.1$~nHz.
These plots are close to being antisymmetric across equator.
The low latitude ($\leq30\degree$) power is stronger close to solar maximum and corresponds to active region flows. 
The power at high latitudes, coming primarily from the $m=1$ HL$+$ mode, is stronger during the quiet Sun period.
Due to the sum over $m$, other modes also contribute to the power at high latitudes, especially the $m=2$ HL$-$ and the $m=3$ ER modes.
The active region flows transport angular momentum poleward, in opposition to the modes.
}
\label{f:rs_unfiltcarr}
\end{figure}

\end{document}